\documentclass[final,notitlepage,12pt]{article}%
\usepackage{epstopdf}
\usepackage{float}
\usepackage{amsfonts}
\usepackage{amssymb}
\usepackage{amsmath}
\usepackage{amsthm}
\usepackage{lscape}
\usepackage{color}
\usepackage[round]{natbib}
\usepackage{nccmath}
\usepackage[doublespacing]{setspace}
\usepackage{tocloft}
\usepackage{pdflscape}
\usepackage{rotating}
\usepackage{amsfonts}
\usepackage{amssymb}
\usepackage{graphicx}
\usepackage{amsmath}
\usepackage{latexsym}
\usepackage{geometry}
\usepackage{multicol}%

\providecommand{\U}[1]{\protect\rule{.1in}{.1in}}
\newtheorem{theorem}{Theorem}[section]

\numberwithin{equation}{section}
\numberwithin{table}{section}
\numberwithin{figure}{section}

\allowdisplaybreaks[4]

\begin{document}
\newcommand\blfootnote[1]{
\begingroup
\renewcommand\thefootnote{}\footnote{#1}
\addtocounter{footnote}{-1}
\endgroup
}

%\newtheorem{corollary}{Corollary}
%\newtheorem{definition}{Definition}
%\newtheorem{lemma}{Lemma}
%\newtheorem{proposition}{Proposition}
%\newtheorem{remark}{Remark}
%\newtheorem{theorem}{Theorem}
%\newtheorem{assumption}{Assumption}

%\numberwithin{corollary}{section}
%\numberwithin{definition}{section}
%\numberwithin{equation}{section}
%\numberwithin{lemma}{section}
%\numberwithin{proposition}{section}
%\numberwithin{remark}{section}
%\numberwithin{theorem}{section}

\allowdisplaybreaks[4]

\begin{titlepage}
\begin{center}
{\large \bf On the Time Trend of COVID-19: A Panel Data Study\footnote{The first author thanks financial support from National Nature Science Foundation of China under grant number: 71671143; the second and the third authors would like to acknowledge the financial support of the Australian Research Council Discovery Grants program under Grant Number: DP200102769.}}
\medskip
\medskip
{
{\sc

Chaohua Dong$^\dagger$ and Jiti Gao$^{\ddagger}$ and Oliver Linton$^{\star}$ and Bin Peng$^{\ddagger}$\blfootnote{{\em $^{\star}$Corresponding Author}:  Oliver Linton, Faculty of Economics, University of Cambridge, Cambridge CB3 9DD, U.K. Email: obl20@cam.ac.uk}

}

\bigskip

$^\dagger$Zhongnan University of Economics and Law\\ $^{\ddagger}$Monash University \\$^{\star}$University of Cambridge}
\bigskip

\today

\bigskip
\bigskip
\bigskip

\begin{abstract}
In this paper, we study the trending behaviour of COVID-19 data at country level, and draw attention to some existing econometric tools which are potentially helpful to understand the trend better in future studies. In our empirical study, we find that European countries overall flatten the curves more effectively compared to the other regions, while Asia \& Oceania also achieve some success, but the situations are not as optimistic elsewhere. Africa and America are still facing serious challenges in terms of managing the spread of the virus, and reducing the death rate, although in Africa the virus spreads slower and has a lower death rate than the other regions. By comparing the performances of different countries, our results incidentally agree with \cite{Chen2020}, though different approaches and models are considered. For example, both works agree that countries such as USA, UK and Italy perform relatively poorly; on the other hand, Australia, China, Japan, Korea, and Singapore perform relatively better.

\medskip


\medskip
\medskip

\end{abstract}
\end{center}

\end{titlepage}

\section{Introduction}\label{Section1}

Words like ``exponential rate" and ``flatten the curve" have been widely cited by all sorts of social media since the outbreak of the pandemic coursed by COVID-19. Since early 2020, governments of the entire world have been frequently updating their policies in order to manage the spread of the virus, and reduce the death rate while constrained by limited medical resources. Understanding the trending behaviour of the pandemic is therefore crucial from the perspective of policy making.

The paper investigates the trending behaviour of COVID-19 data at country level, and draws attention to some existing econometric tools which are potentially helpful in future work. Trend modelling of COVID-19 data is challenging due to the following reasons at least. First, each country shows a dominating deterministic trend, which wipes out other information. Second, the policy of each country has been updated frequently during the pandemic, so analysis using constant parameters may not reflect these impacts properly. Third, time series analysis cannot be conducted for some countries due to small sample, while pooling data together yields a highly unbalanced dataset. In this study, we aim to model the aforementioned challenges, raise some difficulties, and call for studies which can account for these features simultaneously.

Based on our investigation, we find the following econometric literature is particularly useful. Deterministic time trend modelling (such as \citealp{Phillips2007}, \citealp{Robinson} and \citealp{GLP2020}) helps address the first challenge. Time-varying coefficient models which date back to \cite{robinson1989,robinson1991} or works even earlier are useful to address the second challenge. In some recent studies, both \cite{Chen2020} and \cite{Linton2020} conduct time series analysis on COVID-19 data of selected countries for different purposes, while \cite{LMS2020}  forecast infection of COVID-19 using panel data by a Bayesian methodology. They all agree that time-varying coefficients should be adopted to investigate pandemic data. Factor models and relevant data imputation techniques are closely related to the first and third challenges (e.g., \citealp{BaiNg2002, BaiNg2019}, \citealp{SU201784}, and \citealp{SuMiaoJin}). It is noteworthy that \cite{BaiNg2019} and \cite{SuMiaoJin} have worked out that certain types of random missing data can be dealt within the framework of factor analysis effectively.

In our empirical study, we find that European countries overall flatten the curves more effectively compared to the other regions, while Asia \& Oceania also achieve some success, but the situations are not as optimistic elsewhere. Africa and America are still facing serious challenges in terms of managing the spread of the virus, and reducing the death rate, although in Africa the virus spreads slower and has a lower death rate than the other regions. By comparing the performances of different countries, our results incidentally agree with \cite{Chen2020}, though different approaches and models are considered. For example, both works agree that countries such as USA, UK and Italy perform relatively poorly; on the other hand, Australia, China, Japan, Korea, and Singapore perform relatively better.

The rest of this paper is as follows. Section \ref{Section2} presents the model, and the estimation strategy with associated asymptotic properties. In section \ref{Section3}, we provide our empirical findings. Section \ref{Section4} concludes. Theoretical development, tables and figures are provided in the appendix.

Before proceeding further, it is convenient to introduce some notation that will be used throughout this paper. $\lfloor A\rfloor$ means the largest integer not exceeding $A$; $K(\cdot)$ and $h$ represent a kernel function and a bandwidth of the nonparametric kernel method, respectively; $K_{h}( u)  = K( u/h)/h  $; $\mathbb{I}(\cdot)$ stands for the indicator function; $\text{diag}\{A_1, \ldots, A_k\}$ means constructing a diagonal matrix from $A_1, \ldots, A_k$.

\section{Methodology}\label{Section2}

In this section, we consider two models which we believe are useful to investigate the time trend of COVID-19.

\subsection{Model 1}\label{Section2.1}

We now present the first model, which captures the trend aspect. The countries, indexed by $i=1,\ldots, N$, start experiencing the virus at different time points $b_{iT}\in \{1,\ldots, T\}$. For many countries we may have $b_{iT} = 1$, but not all of them. We now propose the following model:

\begin{eqnarray}\label{eq21}
y_{it} =\left\{\begin{array}{ll}
g_i(\tau_t) |t-\beta_{t,b_{iT}}|^a + \varepsilon_{it}, & \text{ for }t\ge b_{iT}\\
0, & \text{otherwise}
\end{array} \right.
\end{eqnarray}
In model \eqref{eq21}, $y_{it}$ is the logarithm of the observed number of new cases (plus one to include days that have zero outcomes). $\varepsilon_{it}$ is an error term capturing information less dominating than the trend. Further assumptions will be imposed on $\varepsilon_{it}$ later to account for potential omitting variable issues, to capture second tier information over time, and to allow for certain types of heterogeneity. Theoretically, $\beta_{t,b_{iT}}$ may be unknown. Practically, $\beta_{t,b_{iT}}$ accounts for the impacts of different starting points, and may have different forms depending on the research questions. A commonly used form of $\beta_{t,b_{iT}}$ may be $\beta_{t,b_{iT}} \equiv b_{iT}-1$. This is not the main focus of the paper, as it does not impact on our empirical study very much. The trend of \eqref{eq21} can be regarded as a common feature of the virus. Specifically, the value of $a$ characterizes the rate of infection or death. Larger $a$ indicates a faster rate. $g_i(\cdot)$ is a function to reflect the change of policy over time for the country $i$, and captures some heterogeneous features across countries.

We can regard \eqref{eq21} as a panel data version of \cite{GLP2020} with an extra moving mean $\beta_{t,b_{iT}}$. This raises a few challenges that are raised in both the main text and the online supplementary appendix. Before proceeding further, we impose a condition to quantify the impacts of missing values. Specifically, suppose that there exist a sequence of fixed points $\{ b_1^*,\ldots, b_N^*\}$ and a known function $\beta^*(\cdot ,\cdot) $ such that

\begin{eqnarray}\label{eq22}
(1). \ &&\max_{i\ge 1}\left|\frac{b_{iT}}{T}-b_i^*\right| =O( T^{-\nu_1} ) , \nonumber \\
(2). \ &&\max_{i\ge 1}\left|\left|\frac{t-\beta_{t,b_{iT}}}{T}\right|^a - \beta^*(\tau_t,b_i^*)\right| \leq C_0 \, T^{-\nu_2},
\end{eqnarray}
where $0\leq C_0<\infty$, $\nu_1$ and $\nu_2$ are fixed constants satisfying that $0<\nu_1\le 1$ and $\nu_2>0$. When $\beta_{t,b_{iT}} \equiv b_{iT}-1$ and $\beta^*(\tau_t, b_i^*) = \left|\tau_t - b_i^*\right|^a$, part (2) of (\ref{eq22}) holds trivially. Without missing values, $b_{iT}$'s and $b_i^*$'s reduce to 1 and 0 respectively. Practically, the values of $b_{iT}$'s and $b_i^*$'s can be controlled by removing a reasonable range of periods from the beginning in order to reduce the impacts of missing values. In practice, one has to find a balance between available sample size and the impact of missing data.

\medskip

We are interested in recovering information under the framework of \eqref{eq21}-\eqref{eq22}. To carry on our analysis, we write \eqref{eq21} in vector form.

\begin{eqnarray}\label{eq23}
Y_t =\mathbb{I}_t  G(\tau_t) + \mathbb{I}_t  \mathcal{E}_{t},
\end{eqnarray}
where $\mathbb{I}_t = \text{diag}\{\mathbb{I}(t\ge b_{1T}),\ldots,\mathbb{I}(t\ge b_{NT}) \} $, $Y_t = (y_{1t} ,\ldots, y_{Nt})'$, $\mathcal{E}_t = (\varepsilon_{1t},\ldots, \varepsilon_{Nt})'$, and
\begin{eqnarray*}
G(\tau_t) =\left( g_{1}(\tau_t)|t- \beta_{t,b_{1T}}|^a, \ldots, g_{N}(\tau_t) |t-\beta_{t,b_{NT}}|^a \right)'.
\end{eqnarray*}
Since $G(\cdot)$ is unknown, we adopt the nonparametric kernel approach, and multiply $K_h^{1/2}(\tau_t-u)$ for both sides of \eqref{eq23}. Given $\tau_t$ in a small neighbour of $u$, we obtain

\begin{eqnarray}\label{eq24}
\frac{G (\tau_t)}{T^a} K_h^{1/2}(\tau_t-u)\approx \mathcal{G}(u),
\end{eqnarray}
where $\mathcal{G}(u) =\left( g_1(u) \beta^*(u, b_1^*)  ,\ldots, g_N(u)  \beta^*(u, b_N^*) \right)'$. Thus, after proper normalization (i.e., $T^a$), \eqref{eq24} is the leading vector when analysing $Y_t K_h^{1/2}(\tau_t-u)$. However, $a$ is unknown, so has to be estimated.

In view of \eqref{eq23}-\eqref{eq24} and motivated by the construction of the Financial Stress Index\footnote{The largest eigenvalue and the associated eigenvector is calculated using 18 weekly data series in order to measure the degree of financial stress in the markets. See St. Louis Fed's website for details. https://fred.stlouisfed.org/series/STLFSI2}, we conduct the principle component analysis on the sample quantity

\begin{eqnarray}\label{eq25}
\Sigma(u) = \frac{1}{NT}\sum_{t =1}^T Y_t Y_t' K_h(\tau_t - u)
\end{eqnarray}
for all $u$.

We briefly explain the intuition below. Note that simple algebra yields

\begin{eqnarray}\label{eq26}
\Sigma(u) & = &\frac{1}{NT}\sum_{t=1}^T \mathbb{I}_t G (\tau_t)G (\tau_t)' \mathbb{I}_t K_h(\tau_t-u)  \nonumber \\
&& + \frac{1}{NT}\sum_{t =1}^T \mathbb{I}_t \mathcal{E}_t\mathcal{E}_t' \mathbb{I}_t K_h(\tau_t-u)  +\text{interaction terms}.
\end{eqnarray}
Loosely speaking, $\frac{1}{NT}\sum_{t=1}^T \mathbb{I}_t G (\tau_t)G (\tau_t)' \mathbb{I}_t K_h(\tau_t-u)  $ of \eqref{eq26} contains a quadratic in the time trend that will dominates the other terms. As a consequence, the largest eigenvalue and the associated eigenvector of $\Sigma(u)$ reflect the information associated with $\frac{1}{NT}\sum_{t=1}^T \mathbb{I}_t G (\tau_t)G (\tau_t)' \mathbb{I}_t K_h(\tau_t-u)  $ only, which allows us to focus on the trending properties of the virus, and ignore the secondary information asymptotically. To explain the intuition using an even simpler example, one may consider conducting an OLS regression for $y_t=\rho\, t+\varepsilon_t$, where as long as $\varepsilon_t$ is not diverging faster than $t$, the information of $\rho$ can always be retrieved.

That said, let $\lambda_u$ and $\ell_u$ be the largest eigenvalue and the corresponding eigenvector of $\Sigma(u)$, and $ \|\ell_u\| =1$. Mathematically, it is written as

\begin{eqnarray}\label{eq27}
\lambda_{u}\ell_u = \Sigma(u)\ell_u.
\end{eqnarray}
Accounting for the unbalancedness of the data, we further define the following set:

\begin{eqnarray*}
\mathbb{C} =\Big \{t \ | \ t=1,\ldots, T, \ \lim_{N\to \infty}\frac{\sharp \mathbb{N}_{\tau_t}}{N} =1 \Big \},
\end{eqnarray*}
where $\mathbb{N}_u = \{ i \ |\ b_i^* \le u-h, 1\le i\le N \} $, and $\sharp \mathbb{N}_u$ represents the cardinality of $\mathbb{N}_u$. Let $ \mathbb{N}_u^c = \{ 1,\ldots, N\}\setminus \mathbb{N}_u$. By construction, $\mathbb{C}$ rules out a set of time periods that we cannot make inference on due to the availability of data. In practice, we may let $\sharp \mathbb{N}_{\tau_t}\ge N -\ln N$, which replaces the limit in the definition of $\mathbb{C}$ as a practical guide to choose $\mathbb{C}$. Alternatively, we can let $\mathbb{C} = \{\max_{i\ge 1}{b_{iT}}-c,\ldots, T\}$ with $c$ being a reasonably small positive integer for feasibility and simplicity.

Finally, the estimator of $a$ is presented as follows.

\begin{eqnarray}\label{eq28}
\widehat{a}= \frac{1}{2\ln T }\cdot \ln \Big\{\frac{1}{\sharp \mathbb{C}}  \sum_{t\in \mathbb{C}} \lambda_{\tau_t}\Big\}.
\end{eqnarray}
Intuitively, $\frac{1}{\sharp \mathbb{C}}  \sum_{t\in \mathbb{C}} \lambda_{\tau_t}$ yields an estimate of $O(T^{2a})$ using \eqref{eq26}, so the logarithm of $\frac{1}{\sharp \mathbb{C}}  \sum_{t\in \mathbb{C}} \lambda_{\tau_t}$ is divided by $2\ln T$ to yield an estimate of $a$.

\medskip

Below, we present our assumptions and give some justifications.

\noindent \textbf{Assumption 1}
\begin{enumerate}

\item Let $K(\cdot)$ be a function defined on $[-1,1]$, $K^{(1)}(w)$ be uniformly bounded on $[-1,1]$, $\int_{-1}^{1}K(w)dw=1$ and $\int_{-1}^{1}|w|K(w)dw<\infty$. Suppose that $h\to 0$ and $Th\to \infty$.

\item
\begin{enumerate}
\item Suppose that $\max_{i\ge 1}\sup_{\tau\in \mathbb{D}}|F_i( \tau)|<\infty$, where $F_i( \tau) =  g_i(\tau) \beta^*(\tau, b_i^*)$ and $\mathbb{D} = [\inf_{t\in \mathbb{C}}\tau_t,1]$. As $w\to 0$, let $\max_{i\ge 1}\sup_{\tau\in  \mathbb{D} } | F_i( \tau+w) - F_i( \tau) | \le c|w |^{\mu} $, where $\mu$ and $c >0$ are fixed constants.

\item There exists a function $\bar{g}(u)$ such that $\sup_{u\in \mathbb{D}}|\frac{1}{N}\mathcal{G}(u)'\mathcal{G}(u)- \bar{g}(u)^2| =O(\phi_{2,N})$, where $\phi_{2,N}\to 0$, $\int_{\mathbb{D}} \bar{g}(u)^2 du=1$, and $\mathcal{G}(\cdot)$ is defined in \eqref{eq24}.
\end{enumerate}

\item Suppose that $\sup_{u\in \mathbb{D}}\frac{1}{NT} \sum_{t=1}^T\mathcal{E}_t'  \mathcal{E}_t K_h(\tau_t-u)=O_P(\delta_T)$, and $\delta_T/T^{2a}\to 0$.
\end{enumerate}

Assumption Assumption 1.1 imposes restrictions on the kernel function and the bandwidth, which are standard in the literature of kernel regression (\citealp{LiRacine}).

In Assumption 1.2.a, the condition on $F_i( \tau)$ requires Lipschitz continuity. It can be further decomposed by putting restrictions on $a$, $\beta^*(\cdot,\cdot)$ and $g_i(\cdot)$'s, but it will lead to quite lengthy notation and development. Assumption 1.2.b imposes an identification restriction. The condition $\int_{\mathbb{D}} \bar{g}(u)^2 du=1$ fixes the location of $\bar{g}(u)$ along $Y$-axis, and it has no impact on the quantities in relative terms that we shall explore in the empirical study.

As the error terms include information less dominating than the trend, all we require in Assumption 1.3 is that the magnitude of the secondary information does not overwhelm the trend presented by the virus, which can be regarded as how we model the omitting variable issues in the current setting.

\medskip

We are now ready to present the asymptotic results associated with our empirical investigation.

\begin{theorem}\label{Theorem1}
Consider the model stated in \eqref{eq21} and \eqref{eq22}. Under Assumption 1, as $(N,T)\to  (\infty,\infty)$,

\begin{enumerate}
\item $ \sup_{u\in  \mathbb{D}}\|\ell_u \ell_u' - P_{\mathcal{G}(u)}\|  =  O_P(\phi_{1,NT}) $, and $ \sup_{u\in  \mathbb{D}} \left| \frac{\lambda_u}{T^{2a}}  -  \frac{\mathcal{G} (u) '\mathcal{G} (u)}{N}\right| =  O_P(\phi_{1,NT}) $;

\item $\widehat{a}-a = O_P( \frac{\phi_{1,NT}+\phi_{2,N}}{\ln T})$,
\end{enumerate}
where $\phi_{1,NT}=\frac{\delta_T^{1/2}}{T^a}+\big\{ \frac{1}{T^{\min\{\nu_1, \nu_2 \}}} +\frac{\sharp \mathbb{N}_u^c}{N}+h^{\mu} \big\}^{1/2}$, and $P_{\mathcal{G}(u)} = \mathcal{G}(u)\{\mathcal{G}(u)'\mathcal{G}(u) \}^{-1}\mathcal{G}(u)'$.

\medskip

\noindent In addition, suppose $b_i^*=0$ for $i\ge 1$.
\begin{enumerate}
\item[3.] For $\forall t , s\in \mathbb{C}$, $R_{ts} -\frac{ \beta^*(\tau_t, 0)^2}{\beta^*(\tau_s, 0)^2 } \cdot \frac{ \|\mathbb{G} (\tau_t)\|^2    }{ \|\mathbb{G} (\tau_s)\|^2 } =  O_P(\phi_{1,NT} )  $;

\item[4.] For $\forall i,j \in \mathbb{N}_u$, $\sup_{u\in \mathbb{D}}\left|Q_{u,ij} -\frac{g_i(u)}{g_j(u)}\right|  =  O_P(\phi_{1,NT} ) $,
\end{enumerate}
where $R_{ts} = \frac{\lambda_{\tau_t} }{\lambda_{\tau_s} }$, $\mathbb{G}(u)=(g_1(u),\ldots, g_N(u))'$, $Q_{u,ij} =\frac{\ell_{u,i}}{\ell_{u,j}}$, and $\ell_{u,i}$ stands for the $i^{th}$ element of $\ell_u$.
\end{theorem}

With a balanced dataset, the terms involving $\nu_1$, $\nu_2$, $\mu$ and $\sharp \mathbb{N}_u^c$ in the above theorem will vanish, and the asymptotic development will be much simplified. Utilizing panel data, the rate of the second result improves the slow rate of Theorem 4.2 of \cite{GLP2020}, wherein a detailed explanation can be found.

The first result explains how the unbalancedness of the data affects the asymptotic results. Also, it implies that we can recover the space spanned by $\mathcal{G}(u) $. Under the conditions $b_i^*=0$ for $i\ge 1$, the result will reduce to  $ \sup_{u\in  \mathbb{D}}\|\ell_u \ell_u' - P_{\mathbb{G}(u)}\|  =  O_P(\phi_{1,NT}) $. It is noteworthy that the condition $b_i^*=0$ for $i\ge 1$ indicates that the missing value is negligible in the asymptotic analysis, which can be controlled by choosing $\mathbb{C}$ in practice.

For the third result, without loss of generality, suppose that $t>s$. Note that there are two ratios involved in $R_{ts}$, i.e., $ \frac{|\beta^*(\tau_t, 0)|}{|\beta^*(\tau_s, 0)|}$ and $\frac{ \|\mathbb{G} (\tau_t)\|}{ \|\mathbb{G} (\tau_s)\|}$. It is not hard to see that the ratio $ \frac{|\beta^*(\tau_t, 0)|}{|\beta^*(\tau_s, 0)|}$ measures the rate associated with the virus, while the ratio $\frac{ \|\mathbb{G} (\tau_t)\|}{ \|\mathbb{G} (\tau_s)\|}$ reflects the efforts that the countries make to flatten the curves. For effective policies, the ratio $R_{ts}$ should be lower than 1.

The fourth result is also about a ratio that provides a way of comparing the effectiveness of two different policies at the same time point. Note that $g_i(\cdot)$'s model the effectiveness of the policies. A lower value of $g_i(\cdot)$ indicates better efforts in terms of flattening the curve. Thus, if $0<\frac{g_i(u)}{g_j(u)}<1$, we may conclude the country $i$ has a more effective policy compared to the country $j$. Otherwise, the country $j$ performs relatively better.

Finally, we comment on how $g_i(\cdot)$'s and $\beta^*(\cdot,\cdot)$ can be recovered. Since $\beta^*(\cdot,\cdot)$ and $g_i(\cdot)$'s exist in the model through a multiplication form, they cannot be individually estimated without further identification restrictions. If one is willing to impose a restriction (such as $\frac{\mathbb{G} (u) '\mathbb{G}  (u)}{N}=1$), then $\beta^*(\cdot,\cdot)$ can be recovered as suggested by the second argument of Theorem \ref{Theorem1}.1. If the form of $\beta^*(\cdot,\cdot)$ was known, the asymptotic distribution associated with the estimate of each $g_i(\cdot)$ can be constructed as in Theorem 4.3 of \cite{GLP2020}. Alternatively, Theorem \ref{Theorem1}.4 suggests that for any given $u$ we may pick an individual $i$ as a benchmark, then recover the rest $g_j(u)$'s and $\beta^*(\cdot,\cdot)$ utilizing the ratio of the fourth result. As these are not the main focus of this paper, we leave the choice of identification strategy to future study. In our empirical work we will emphasise the identified quantities: a, and the ratios $R*_{ts}=\frac{ \beta^*(\tau_t, 0)^2}{\beta^*(\tau_s, 0)^2 } \cdot \frac{ \|\mathbb{G} (\tau_t)\|^2    }{ \|\mathbb{G} (\tau_s)\|^2 }$ and $Q(u)=\frac{g_i(u)}{g_j(u)}$.

\subsection{Model 2}

We consider a second model that is designed to capture a single peaked epidemic trajectory, similar to \cite{Linton2020}. We consider the following regression

\begin{eqnarray}\label{eq29}
y_{it} =\left\{\begin{array}{ll}
\gamma_i - g_i(\tau_t) |t-\beta_{t,b_{iT}} |^a + \varepsilon_{it}, & \text{ for }t\ge b_{iT}\\
0, & \text{otherwise}
\end{array} \right.,
\end{eqnarray}
where $\gamma_i$ is the global maximum of each individual. When $t=\beta_{t,b_{iT}}$, the global
maximum is achieved at $\gamma_i$.

If we have a complete trajectory of the epidemic, or at least data that includes the peak and sometime afterwards, we may estimate $\gamma_i$ directly. Specifically, we may take any local (in time) smoother
and maximize this over time. The smoothing method eliminates the error term and then the resulting function is uniquely maximized at the true peak time.  One then can use the methodology of Section \ref{Section2.1} to work with the transformed model as follows.

\begin{eqnarray}\label{eq211}
y_{it}^* =\left\{\begin{array}{ll}
g_i(\tau_t) |t-\beta_{t,b_{iT}} |^a + \varepsilon_{it}^*, & \text{ for }t\ge b_{iT}\\
0, & \text{otherwise}
\end{array} \right.,
\end{eqnarray}
where $y_{it}^* =\widehat{\gamma}_i -y_{it}$ and $\varepsilon_{it}^* =-\varepsilon_{it} + (\widehat{\gamma}_i-\gamma_i )$.

Additionally, one may consider an estimation strategy that tries to estimated the parameters of interest simultaneously to avoid the bias caused by the plug-in procedure. We wish to leave it to the future study, but we examine the model \eqref{eq211} using the approach of Section \ref{Section2.1} in the empirical study as a robustness check.

\section{Empirical Study}\label{Section3}

In this section, we investigate the time trend of the COVID-19 data. Before proceeding further, we comment on two practical issues --- the choice of kernel function and the bandwidth selection procedure.

For the kernel function, we follow \cite{HongLi} and \cite{SU201784} to adopt a boundary adjusted kernel:

\begin{eqnarray*}
K((\tau_t -u)/h) =\left\{ \begin{array}{ll}
\mathcal{K}((\tau_t - u)/h) , & u\in [h,1-h] \\
\mathcal{K}((\tau_t -u)/h) /\int_{-1}^{(1-u)/h} \mathcal{K}(w)dw, & u\in (1-h,1]
\end{array}\right.
\end{eqnarray*}
for $t=1,\ldots, T$, where $\mathcal{K}(w)$ is the Epanechnikov kernel. By construction of $\mathbb{C}$, there is no need to adjust the left boundary.

Next, we provide a bandwidth selection procedure which minimizes a leave-one-out cross validation function as follows.

\begin{eqnarray*}
\widehat{h} = \text{argmin}_h \text{CV}(h),
\end{eqnarray*}
and

\begin{eqnarray*}
\text{CV}(h) = \sum_{t\in \mathbb{C}}\| Y_t/(\sqrt{N}T^{a_h}) -\widehat{\ell}_{-\tau_t}\|^2,
\end{eqnarray*}
where $a_h$ is obtained from \eqref{eq28} given $h$, and $\widehat{\ell}_{-\tau_t}$ is obtained from \eqref{eq27} by replacing $\Sigma(\tau_t)$ with $ \frac{1}{NT}\sum_{s =1,s\ne t}^T Y_s Y_s' K_h(\tau_s - \tau_t)$. The terms $\sqrt{N} $ and $T^{a_h}$ are normalizers to ensure that $\widehat{\ell}_{-\tau_t}$ and the normalized $Y_t$ are on the same scale. To examine the sensitivity of the bandwidth selection procedure, we further consider $h_L = 0.8\widehat{h}$ and $h_R =1.2\widehat{h}$.

\subsection{Data}\label{Section3.1}

We focus on daily new infection and new deaths from four regions\footnote{The data are downloaded from European Centre for Disease Prevention and Control: https://www.ecdc.europa.eu/en/publications-data/download-todays-data-geographic-distribution-covid-19-cases-worldwide.}, (i.e., Africa (AF), America (AM), Asia \& Oceania (AO), and Europe (EU)); we account for population density of each country in the following analysis. Note that there are only 8 countries from Oceania in the data source, so we merge Asia and Oceania together. Population density is based on the data of 2018 from World Bank, and is measured as people per sq. km of land area. We exclude countries that do not have the population density figures. For each region, the sample period starts from the date when the first confirmed case is recorded, but we remove the first 30 days of each region in order to reduce the impacts of missing data. Finally, we summarize the available sample in Table \ref{Table1}.

For infection data, the four regions have roughly the same number of countries. However, death data are very unbalanced. We remove the countries with total deaths less than 20 at 31/05/2020, which is why the number of countries drops for death data.  It is not surprising that Asia \& Oceania has the longest period due to early outbreak of China, while Africa has the shortest period.

\subsection{Results Associated with Model \eqref{eq21}}\label{Section3.2}

We now start conducting numerical analysis using the approach of Section \ref{Section2.1}. Specifically, we consider two sets of $\{y_{it}\}$ for both infection and death.

\begin{enumerate}
\item[] Case 1: $\ln\left(\text{daily increase}+1\right)$
\item[] Case 2: $\ln\left(\frac{\text{daily increase } +1}{\text{population density}}\right)$
\end{enumerate}

\subsubsection{Overall Analysis}\label{Section3.2.1}

We let $\mathbb{C} = \{ \lfloor T/4\rfloor + 1,\ldots, T\}$ for simplicity, and summarize the estimates of $a$ in Table \ref{Table2}, which shows that the estimates are not overly sensitive to different choices of the bandwidth.

For infection data, Europe has the highest values of $\widehat{a}$ for both Cases 1 and 2, which could be due to overall high quality infrastructure leading to high mobility of the entire population. Moreover, America and Asia \& Oceania have roughly similar values in both Cases 1 and 2, while Africa has the lowest value, which implies that the virus spreads in Africa slower than the other regions.

For death data, America has the highest death rate for both Cases 1 and 2. Although the estimates from the original data (i.e., Case 1) indicate that Africa has a very low death rate, the estimates from the normalized version (i.e., Case 2) indicates that the situation is not too optimistic but is still the best among four regions.

\medskip

Next, we examine the ratio $R_{t+1, t} $ for $t= \lfloor T/4 \rfloor+1,\ldots, T-1$ by the third result of Theorem \ref{Theorem1}, and plot them in Figures \ref{Infection_R} and \ref{Death_R} for infection and death data respectively. We explore infection in Figure \ref{Infection_R} first. For Case 1, the curves of Africa and America are always above 1, although approaching to 1 slowly. However, for Case 2, the plot of America diverges from 1 during the entire period. Europe is the only region achieving a rate lower than 1 during April and most time of May based on the original data and normalized data. In this sense, we believe the policies of European countries are most effective. The curves of Asian \& Oceania move around 1 all the time for both cases, but is slightly higher than 1 in most of the days. It is noteworthy that most curves of Figure \ref{Infection_R} diverging from 1 from late May, which might be due to the fact that many governments start lifting the lock-down in May. For the ratios associated with death data in Figure \ref{Death_R}, the patterns are almost identical to those presented in Figure \ref{Infection_R}, so we do not repeat the discussions.

\subsubsection{Comparison across Countries}\label{Section3.2.2}

In this subsection, we compare the performances of countries in each region using the fourth result of Theorem \ref{Theorem1}. Specifically, for each region, we let the country that has the largest value of daily increase at 31/05/2020 be the benchmark, and label it by the index $i=1$. We summarize the reference countries in Table \ref{Table3}. We then plot $Q_{\tau_t,i1}$ for $i\ge 2$ associated with infection and death in Figures \ref{Infection_Q} and \ref{Death_Q} respectively, where the countries are labelled by ISO 3166-1 alpha-3 codes. In each sub-plot, the legend is ranked by $Q_{\tau_t,i1}$ from largest to smallest at the time period $T$. The lines in each sub-plot reflect how the corresponding countries perform at different time points compared to the reference country. As explained under Theorem \ref{Theorem1}, (1). smaller value indicates better performance, and (2). a value less (greater) than 1 indicates better (worse) performance than the reference country.

Our results somewhat agree with the findings of \cite{Chen2020}. For example, (1). countries such as USA, UK and Italy are at the top of the corresponding sub-plots in our investigation, which indicates ineffective performance in terms of managing the spread of the virus and reducing the death rate; (2). on the other hand, our finding also suggests that countries such as Australia, China, Japan, Korea, and Singapore perform relatively well as in \cite{Chen2020}.

\subsubsection{Rolling-Window Analysis}\label{Section3.2.3}

Finally, we estimate $a$ and the ratio $R$ using a rolling-window sample in order to capture some dynamics, which in a sense can be regarded as a robustness check on the sensitivity of the data. We prepare the data as in Section \ref{Section3.1}, and remove the first 40 days for each region to avoid the impacts of missing value on the 30 days rolling-window (i.e., $T=30$ for each regression). For each window, we let $\mathbb{C}=\{ 26,27,\ldots, 30\}$ and estimate $\bar{R} = \frac{1}{4}\sum_{t=26}^{29}R_{t+1,t}$. We then record the estimated $a$ and $\bar{R}$ from the first available window till the end.

For effective policies, we expect the estimates of $a$ show a turning point at certain stage, and expect the value of $\bar{R}$ below one. We plot the estimates of each region in Figures \ref{RW_a_Inf}-\ref{RW_R_Dea}, where the $X$-axis is indexed by the last day of the consecutive 30 days period.

First, we take a look at the values associated with infection in Figures \ref{RW_a_Inf} and \ref{RW_R_Inf}. In Figure \ref{RW_a_Inf}, the curves of Africa and America keep increasing with a very steady rate, which is a concern from the perspective of flattening the curve. The curves of Asia \& Oceania become flat gradually, but the turning points have not shown up yet. Europe is the only continent which has a turning point in Figure \ref{RW_a_Inf}, and the pattern exists in both Cases 1 and 2. It further supports that European countries have more effective polices overall. In Figure \ref{RW_R_Inf}, the curves of Asia \& Oceania and Europe are approaching to 1, while the curves of Africa and America do not. Especially, the values of $\bar{R}$ of Africa start diverging from 1 from late May, which is also worrisome.

Second, we turn to the results associated with death in Figures \ref{RW_a_Dea} and \ref{RW_R_Dea}. Clearly, in Figure \ref{RW_a_Dea}, the death rate of Europe has been dropping, while Asia \& Oceania have managed to flatten the curve, but the turning point has not shown up yet. Africa and America have increasing death rates during the entire period. In Figure \ref{RW_R_Dea}, Europe still performs much better than the other regions, as it is the only region having $\bar{R}$ less than 1. The curves of Asia \& Oceania have been approaching to 1, while Africa and America do not show much improvement during the period.

\subsection{Results Associated with Model \eqref{eq29}}

The data and the corresponding settings of this subsection are identical to those in Section \ref{Section3.2}, but we work with the transferred version using \eqref{eq211}. Still, we consider Cases 1 and 2 for the transferred data. It is noteworthy that under the model \eqref{eq29}, the interpretation on the values of $a$, $R_{t+1, t} $ and $Q_{\tau_t, i1}$ are respectively different from those in Section \ref{Section3.1}. Specifically, the effective policies would ensure relatively short periods to reach the peak of the pandemic. In this sense, the first different is that large $a$ may not be a sign of bad situation. The second difference is that we expect the ratio $R_{t+1, t} $ greater than 1 to indicate more effective policies, since larger $R_{t+1, t} $ implies reaching the peak with a shorter period. Finally, for the ratio $Q_{\tau_t, i1}$ with  $i=2,\ldots, N$, we expect a value greater than 1 to represent a more effective policy compared to the reference individual.

Note that since Africa and America have not reached the peak with obvious reasons by screening the data plots, we do not comment on the values associated with Africa and America much below although the values for these two regions are reported.

\subsubsection{Overall Analysis}

We first summarize the estimates of $a$ in Table \ref{Table4}. For both infection and death data, it seems to suggest that in Europe the spread of the virus and the death rate reach the peak slower than the other regions by nature. For Asian \& Oceania, the spread of the virus and the death rate tend to reach the peak slightly faster than Europe for both Cases 1 and 2.

We now focus on the values of $R_{t+1, t} $ presented in Figures \ref{Infection_R_Extension} and \ref{Death_R_Extension}. Consistent with what we find in Section \ref{Section3.2.1}, Europe indeed has more effective polices, as the values of $R_{t+1, t} $ are greater than 1 in the entire period for both Cases 1 and 2. Asia \& Oceania have some success, but the situation is not as good as in Europe.

\subsubsection{Comparison across Countries}

For each region, the reference countries are the same as those in Table \ref{Table3}. The legend of each sub-plot is ranked by $Q_{\tau_t,i1}$ from largest to smallest at the time period $T$, however, larger value implies better performance in this case.

For the infection data of Europe, Case 1 of Figure \ref{Infection_Q_Extension} fully agrees with Case 1 of Figure \ref{Infection_Q}, i.e., all countries perform better than the reference country. For the death data of Europe, a similar argument applies to Case 2 of Figure \ref{Death_Q_Extension} and Case 2 of Figure \ref{Death_Q}.

Interestingly, for Asia \& Oceania, the downward trending of Cases 1 and 2 in Figures \ref{Infection_Q} and \ref{Death_Q} becomes upward trending in Figures \ref{Infection_Q_Extension} and \ref{Infection_R_Extension}. Thus, both models confirm that compared to the reference country, the rest countries in Asia \& Oceania have been improving, or the situation of the reference country has been getting out of control.

\section{Conclusion}\label{Section4}

In this paper, we study the trending behaviour of COVID-19 data at country level, and draw attention to some existing econometric tools which are potentially helpful to understand the trend better in the future study. In our empirical study, we find that European countries overall flatten the curves more effectively compared to the other regions, while Asia \& Oceania also achieve some success, but the situations are not optimistic as in Europe. Africa and America are still facing serious challenges in terms of managing the spread of the virus, and reducing the death rate, although in Africa the virus spreads slower and has lower death rate than the other regions by nature. By comparing the performances of different countries, our results incidentally agree with \cite{Chen2020}, though different approaches and models are considered. For example, both works agree that countries such as USA, UK and Italy perform relatively poorly; on the other hand, Australia, China, Japan, Korea, and Singapore perform relatively better.

\bigskip

{\small

\bibliography{Refs}

% ***********************************************************************

\renewcommand{\theequation}{A.\arabic{equation}}
\renewcommand{\thesection}{A.\arabic{section}}
\renewcommand{\thefigure}{A.\arabic{figure}}
\renewcommand{\thetable}{A.\arabic{table}}
%\renewcommand{\thelemma}{A.\arabic{lemma}}

% redefine the command that creates the equation no.
% reset counter
\setcounter{equation}{0}
\setcounter{section}{0}
\setcounter{table}{0}
\setcounter{figure}{0}

\bigskip

\section*{Appendix A}

In what follows, $O(1)$ stands for a constant, and may be different at each appearance. Without loss of generality, let $ b_{1T}\le b_{2T}\le \cdots \le b_{NT}$ in what follows.

\medskip

\noindent \textbf{Lemma A.1}

{\it
Consider the model stated in \eqref{eq21} and \eqref{eq22}. Under Assumption 1, as $(N,T)\to  (\infty,\infty)$,

\begin{enumerate}
\item  $\sup_{u\in  \mathbb{D}}\left\| \frac{1}{NT}\sum_{t=1}^T \mathbb{I}_t  \frac{G (\tau_t)}{T^a} \frac{G (\tau_t)'}{T^a} \mathbb{I}_t K_h(\tau_t-u) -  \frac{1}{N}\mathcal{G}(u)\mathcal{G}(u)' \right\|^2\\ = O\left(\frac{1}{T^{\min\{\nu_1, \nu_2 \}}}+\frac{\sharp \mathbb{N}_u^c}{N} +h^{\mu}\right) $;

\item $\sup_{u\in \mathbb{D}}\left|\frac{1}{NT} \sum_{t=1}^T G (\tau_t)'\mathbb{I}_t  G (\tau_t)\left(\beta^*(\tau_t, b_i^*)\right)^{2} K_h(\tau_t-u) -\frac{1}{N} \sum_{i\in \mathbb{N}_u} \beta^*(u,b_i^*)^2 g_{i}(u)^2\right|\\
= O\left(\frac{1}{T^{\min\{\nu_1, \nu_2 \}}}+\frac{\sharp \mathbb{N}_u^c}{N} +h^{\mu}\right)$.
\end{enumerate}
}

\noindent \textbf{Proof of Lemma A.1:}

(1). Write

\begin{eqnarray*}
&&\sup_{u\in  \mathbb{D}}\left\| \frac{1}{NT}\sum_{t=1}^T \mathbb{I}_t  \frac{G (\tau_t)}{T^a} \frac{G (\tau_t)'}{T^a} \mathbb{I}_t K_h(\tau_t-u) -  \frac{1}{N}\mathcal{G}(u)\mathcal{G}(u)' \right\|^2  \\
&=&\sup_{u\in  \mathbb{D}}\frac{1}{N^2}\sum_{i\in \mathbb{N}_u} \sum_{j\in \mathbb{N}_u}\Big\{\frac{1}{T}\sum_{t=1}^T F_i(u )F_j(u )  \mathbb{I}(t\ge b_{iT})  \mathbb{I}(t\ge b_{jT})  K_h(\tau_t-u)  -F_i(u )F_j(u )\Big\}^2\\
&&+O\left(\frac{1}{T^{\nu_2}}+\frac{\sharp \mathbb{N}_u^c}{N}\right) \\
&=&\sup_{u\in  \mathbb{D}}\frac{1}{N^2}\sum_{i\in \mathbb{N}_u} \Big\{\int_{b_i^*}^1 F_i(u )^2 K_h(w-u) dw-F_i(u )^2 \Big\}^2 \\
&&+\sup_{u\in  \mathbb{D}}\frac{2}{N^2}\sum_{i\in \mathbb{N}_u} \sum_{j\in \mathbb{N}_u,j<i}\Big\{\int_{b_i^*}^1 F_i(u )F_j(u )  K_h(w-u)dw - F_i(u )F_j(u )\Big\}^2\\
&&+O\left(\frac{1}{T^{ \min\{\nu_1, \nu_2 \}}}+\frac{\sharp \mathbb{N}_u^c}{N} \right) \\
&=&\sup_{u\in  \mathbb{D}}\frac{2}{N^2}\sum_{i\in \mathbb{N}_u} \sum_{j\in \mathbb{N}_u,j<i}\Big\{\int_{b_i^*}^1 F_i(u )F_j(u )  K_h(w-u)dw - F_i(u )F_j(u )\Big\}^2\\
&&+O\left(\frac{1}{T^{ \min\{\nu_1, \nu_2 \}}}+\frac{\sharp \mathbb{N}_u^c}{N} \right) \\
&=& \sup_{u\in  \mathbb{D}}\frac{2}{N^2}\sum_{i\in \mathbb{N}_u} \sum_{j\in \mathbb{N}_u,j<i}\Big\{\int_{-1}^1 F_i(u+hw)F_j(u+hw)  K(w) dw  -F_i(u )F_j(u ) \Big\}^2\\
&&+O\left(\frac{1}{T^{ \min\{\nu_1, \nu_2 \}}}+\frac{\sharp \mathbb{N}_u^c}{N} \right) \\
&=& O\left(\frac{1}{T^{\min\{\nu_1, \nu_2 \}}}+\frac{\sharp \mathbb{N}_u^c}{N} +h^{\mu}\right) ,
\end{eqnarray*}
where the first equality follows from the second condition of \eqref{eq22} and the construction of $\mathbb{C}$; the second equality follows from the definition of Riemann integral and the first condition of \eqref{eq22}; the third equality follows from by focusing on the leading term only; the fourth equality follows by the construction of $\mathbb{C}$ and integration by substitution; and the fifth equality follows from Assumption 1.2.

\medskip

(2). Similar to the first result, the second result follows.  The proof is complete now. \hspace*{\fill}{$\blacksquare$}

\bigskip

\noindent \textbf{Proof of Theorem \ref{Theorem1}:}

(1). We expand \eqref{eq27} as follows.

\begin{eqnarray}\label{exp1}
\frac{\lambda_u}{T^{2a}}\ell_u = \frac{1}{T^{2a}} \Sigma(u) \ell_u &=& \frac{1}{NT^{2a+1}}\sum_{t=1}^T \mathbb{I}_t G (\tau_t)G (\tau_t)' \mathbb{I}_t K_h(\tau_t-u)\ell_u \nonumber \\
&&+ \frac{1}{NT^{2a+1}}\sum_{t =1}^T \mathbb{I}_t \mathcal{E}_t\mathcal{E}_t' \mathbb{I}_t K_h(\tau_t-u) \ell_u \nonumber \\
&& + \frac{1}{NT^{2a+1}}\sum_{t =1}^T \mathbb{I}_t G (\tau_t) \mathcal{E}_t' \mathbb{I}_t K_h(\tau_t-u) \ell_u \nonumber \\
&& +\frac{1}{NT^{2a+1}}\sum_{t =1}^T \mathbb{I}_t \mathcal{E}_t G (\tau_t)' \mathbb{I}_t K_h(\tau_t-u) \ell_u\nonumber \\
&:=& A_1+\cdots + A_4.
\end{eqnarray}
Below, we investigate $A_1$ to $A_4$ one by one.

\begin{eqnarray*}
&&\sup_{u\in  \mathbb{D}} \|A_2 \| =\sup_{u\in  \mathbb{D}} \left\| \frac{1}{NT^{2a+1}}\sum_{t =1}^T \mathbb{I}_t \mathcal{E}_t \mathcal{E}_t' \mathbb{I}_t  K_h(\tau_t-u) \ell_u\right\| \\
&\le & \frac{1}{T^{2a}}\sup_{u\in  \mathbb{D}}\left\{ \frac{1}{NT} \sum_{t=1}^T\mathcal{E}_t' \mathbb{I}_t \mathcal{E}_t K_h(\tau_t-u) \right\}^{1/2}  \left\{ \frac{1}{NT} \sum_{t=1}^T\mathcal{E}_t' \mathbb{I}_t \mathcal{E}_t K_h(\tau_t-u) \right\}^{1/2} \\
&=&O_P\left(1\right)  \frac{1}{T^{2a}}\cdot \delta_T^{1/2} \cdot \delta_T^{1/2} =O_P\left(\frac{\delta_T }{T^{2a}}\right) ,
\end{eqnarray*}
where the first inequality follows from Cauchy-Schwarz inequality; the second equality follows from Lemma A.1 and Assumption 1.3

For $A_3$, write

\begin{eqnarray*}
&&\sup_{u\in  \mathbb{D}} \|A_3 \| =\sup_{u\in  \mathbb{D}} \left\| \frac{1}{NT^{2a+1}}\sum_{t =1}^T \mathbb{I}_t G (\tau_t) \mathcal{E}_t' \mathbb{I}_t  K_h(\tau_t-u) \ell_u\right\| \\
&\le & \frac{1}{T^a}\sup_{u\in  \mathbb{D}}\left\{ \frac{1}{NT} \sum_{t=1}^T \frac{G (\tau_t)'}{T^a}\mathbb{I}_t  \frac{G (\tau_t)}{T^a}  K_h(\tau_t-u) \right\}^{1/2} \left\{ \frac{1}{NT} \sum_{t=1}^T\mathcal{E}_t' \mathbb{I}_t \mathcal{E}_t K_h(\tau_t-u) \right\}^{1/2} \\
&=&O_P\left(1\right)  \frac{1}{T^a}\cdot 1\cdot \delta_T^{1/2} =O_P\left(\frac{\delta_T^{1/2}}{T^a}\right) ,
\end{eqnarray*}
where the first inequality follows from Cauchy-Schwarz inequality; the second equality follows from Lemma A.1 and Assumption 1.3. Similarly,

\begin{eqnarray*}
\sup_{u\in  \mathbb{D}}\|A_4\| =O_P\left(\frac{\delta_T^{1/2}}{T^a}\right).
\end{eqnarray*}
Thus, we obtain that

\begin{eqnarray*}
\sup_{u\in  \mathbb{D}}\left\|\frac{\lambda_u}{T^{2a}} \ell_u - A_1\right\| &=&\sup_{u\in  \mathbb{D}}\left\|\frac{\lambda_u}{T^{2a}}\ell_u - \frac{1}{N}\mathcal{G}(u)\mathcal{G}(u)'\ell_u + \frac{1}{N}\mathcal{G}(u)\mathcal{G}(u)'\ell_u -A_1\right\|\\
& \le &\sum_{j=2}^{4}  \sup_{u\in  \mathbb{D}}\left\| A_j \right\|= O_P\left(\frac{\delta_T^{1/2}}{T^a}\right),
\end{eqnarray*}
which, in connection with Lemma A.1, yields that

\begin{eqnarray}\label{factor1}
\sup_{u\in  \mathbb{D}}\left\|\frac{\lambda_u}{T^{2a}}\ell_u - \frac{1}{N}\mathcal{G}(u)\mathcal{G}(u)'\ell_u  \right\| =   O_P (\phi_{1,NT} ),
\end{eqnarray}
where $\phi_{1,NT}=\frac{\delta_T^{1/2}}{T^a}+\big\{ \frac{1}{T^{\min\{\nu_1, \nu_2 \}}} +\frac{\sharp \mathbb{N}_u^c}{N}+h^{\mu} \big\}^{1/2}$. Thus, the first argument $ \sup_{u\in  \mathbb{D}}\|\ell_u \ell_u' - P_{\mathcal{G}(u)}\|  =  O_P(\phi_{1,NT}) $ follows.

Left multiplying \eqref{factor1} by$\frac{\mathcal{G}(u)'}{\sqrt{N}}$, we can write

\begin{eqnarray*}
\sup_{u\in  \mathbb{D}}\left| \frac{\lambda_u}{T^{2a}}\frac{\mathcal{G} (u)'\ell_u}{\sqrt{N}} - \frac{\mathcal{G} (u) '\mathcal{G} (u)}{N} \cdot \frac{\mathcal{G} (u)'\ell_u}{\sqrt{N}} \right|=  O_P ( \phi_{1,NT} ),
\end{eqnarray*}
which further leads to

\begin{eqnarray}
\sup_{u\in  \mathbb{D}} \left| \frac{\lambda_u}{T^{2a}}  -  \frac{\mathcal{G} (u) '\mathcal{G} (u)}{N}\right| =  O_P(\phi_{1,NT}) .\label{eqA.3}
\end{eqnarray}
Thus, the proof of the first result is complete.

\medskip

(2). Write

\begin{eqnarray*}
\frac{1}{\sharp \mathbb{C}}\sum_{t\in \mathbb{C}}\frac{\lambda_{\tau_t}}{T^{2a}} &=& \frac{1}{ \sharp \mathbb{C}}\sum_{t\in \mathbb{C}}\left( \frac{\lambda_{\tau_t}}{T^{2a}} - \frac{\mathcal{G} (\tau_t) ' \mathcal{G} (\tau_t)}{N}\right)+\frac{1}{\sharp \mathbb{C}}\sum_{t\in \mathbb{C}} \frac{\mathcal{G} (\tau_t)' \mathcal{G} (\tau_t)}{N} \nonumber\\
&=& \frac{1}{\sharp \mathbb{C}}\sum_{t\in \mathbb{C}} \frac{\mathcal{G} (\tau_t)' \mathcal{G} (\tau_t)}{N} +O_P(\phi_{1,NT})\nonumber\\
&=& \int_{\mathbb{D}} \bar{g}(u)^2du +O_P(\phi_{1,NT}+\phi_{2,N})= 1 +O_P(\phi_{1,NT}+\phi_{2,N}),
\end{eqnarray*}
where the second equality follows from \eqref{eqA.3} and the construction of $\mathbb{C}$; and the third equality follows from Assumption 1.2. Then

\begin{eqnarray*}
\ln \Big\{\frac{1}{\sharp \mathbb{C}}\sum_{t\in \mathbb{C}}\lambda_{\tau_t}\Big\} -2a\ln T =O_P(\phi_{1,NT}+\phi_{2,N}),
\end{eqnarray*}
which gives that $\widehat{a}-a = O_P( \frac{\phi_{1,NT}+\phi_{2,N}}{\ln T})$. Thus, the second result follows.

\medskip

(3)-(4). By the first result, the third and fourth results follow immediately. The proof is complete. \hspace*{\fill}{$\blacksquare$}

\newpage

\begin{table}[h]
\centering
\small
\caption{Available Sample}\label{Table1}
\begin{tabular}{lrrrrr}
\hline \hline
 & \multicolumn{2}{c}{Infection} & \multicolumn{1}{l}{} & \multicolumn{2}{c}{Death} \\
 & $N$ & $T$ &  & $N$ & $T$ \\ \cline{2-3} \cline{5-6}
AF & 48 & 62 &  & 23 & 50 \\
AM & 43 & 98 &  & 20 & 61 \\
AO & 48 & 123 &  & 25 & 108 \\
EU & 49 & 98 &  & 41 & 69 \\
\hline \hline
\end{tabular}
\end{table}

\begin{table}[h]
\centering
\small
\caption{Estimate of $a$ for Model \eqref{eq21}}\label{Table2}
\begin{tabular}{llrrrrrrr}
\hline \hline
 &  & \multicolumn{3}{c}{Case 1} & \multicolumn{1}{l}{} & \multicolumn{3}{c}{Case 2} \\
 &  & $\widehat{h}$ & $h_L$ & $h_R$ &  & $\widehat{h}$ & $h_L$ & $h_R$ \\ \cline{3-5} \cline{7-9}
Infection & AF & 0.239 & 0.240 & 0.239 &  & 0.393 & 0.393 & 0.393 \\
 & AM & 0.274 & 0.277 & 0.272 &  & 0.402 & 0.402 & 0.402 \\
 & AO & 0.262 & 0.263 & 0.262 &  & 0.409 & 0.401 & 0.405 \\
 & EU & 0.328 & 0.329 & 0.327 &  & 0.449 & 0.449 & 0.448 \\
 &  &  &  &  &  &  &  &  \\
Death & AF & 0.021 & 0.022 & 0.020 &  & 0.321 & 0.322 & 0.321 \\
 & AM & 0.285 & 0.286 & 0.285 &  & 0.445 & 0.445 & 0.445 \\
 & AO & 0.133 & 0.133 & 0.132 &  & 0.367 & 0.366 & 0.365 \\
 & EU & 0.235 & 0.235 & 0.234 &  & 0.409 & 0.401 & 0.405 \\
 \hline \hline
\end{tabular}
\end{table}

\begin{table}[h]
\centering
\caption{Reference Countries for Figures \ref{Infection_Q}, \ref{Death_Q}, \ref{Infection_Q_Extension}, \ref{Death_Q_Extension}}\label{Table3}
\begin{tabular}{lrr}
\hline\hline
 & Infection & Death \\ \cline{2-3}
AF & South Africa & Egypt \\
AM & Brazil & Brazil \\
AO & India & India \\
EU & Russia & United Kingdom \\
\hline\hline
\end{tabular}
\end{table}

\begin{table}[h]
\centering
\small
\caption{Estimate of $a$ for Model \eqref{eq29}}\label{Table4}
\begin{tabular}{llrrrrrrr}
\hline \hline
 &  & \multicolumn{3}{c}{Case 1} & \multicolumn{1}{l}{} & \multicolumn{3}{c}{Case 2} \\
 &  & $\widehat{h}$ & $h_L$ & $h_R$ &  & $\widehat{h}$ & $h_L$ & $h_R$ \\ \cline{3-5} \cline{7-9}
 Infection & AF & 0.229 & 0.230 & 0.229 &  & 0.396 & 0.396 & 0.396 \\
 & AM & 0.187 & 0.188 & 0.186 &  & 0.358 & 0.358 & 0.357 \\
 & AO & 0.210 & 0.212 & 0.209 &  & 0.393 & 0.394 & 0.392 \\
 & EU & 0.164 & 0.164 & 0.164 &  & 0.375 & 0.375 & 0.375 \\
 &  &  &  &  &  &  &  &  \\
Death & AF & 0.108 & 0.109 & 0.107 &  & 0.339 & 0.339 & 0.339 \\
 & AM & 0.096 & 0.096 & 0.095 &  & 0.367 & 0.367 & 0.367 \\
 & AO & 0.144 & 0.146 & 0.143 &  & 0.373 & 0.373 & 0.372 \\
 & EU & 0.103 & 0.104 & 0.102 &  & 0.345 & 0.345 & 0.344 \\
 \hline \hline
\end{tabular}
\end{table}

\newpage

\begin{figure}[h]\caption{Model 1 --- $R_{t+1,t}$ of Infection Data. The left and right panels are Case 1 and Case 2 respectively. }\label{Infection_R}
\centering
\hspace*{-1cm} \includegraphics[scale=0.5]{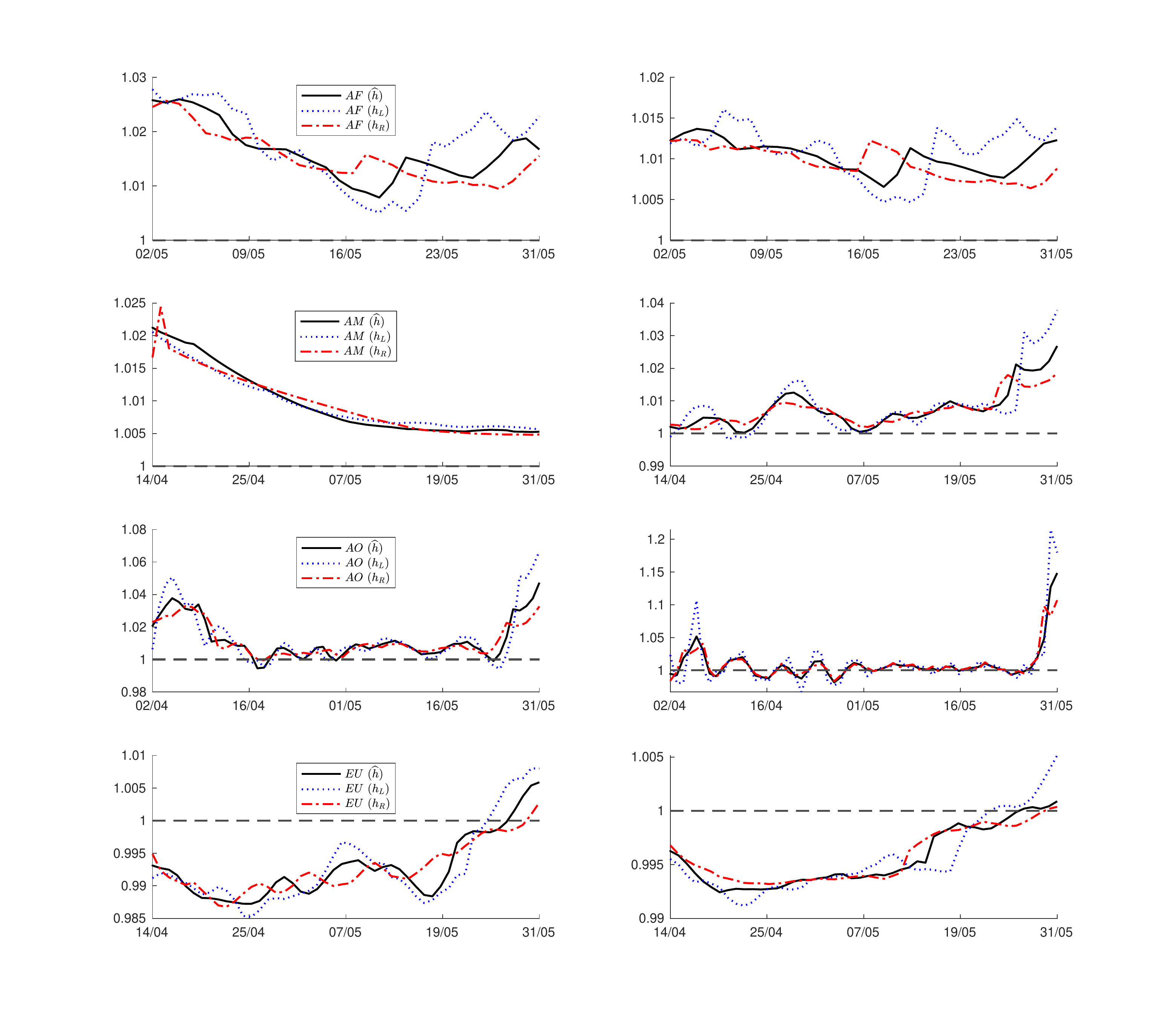} \\
\end{figure}

\begin{figure}[h]\caption{Model 1 --- $R_{t+1,t}$ of Death Data. The left and right panels are Case 1 and Case 2 respectively.}\label{Death_R}
\centering
\hspace*{-1cm} \includegraphics[scale=0.5]{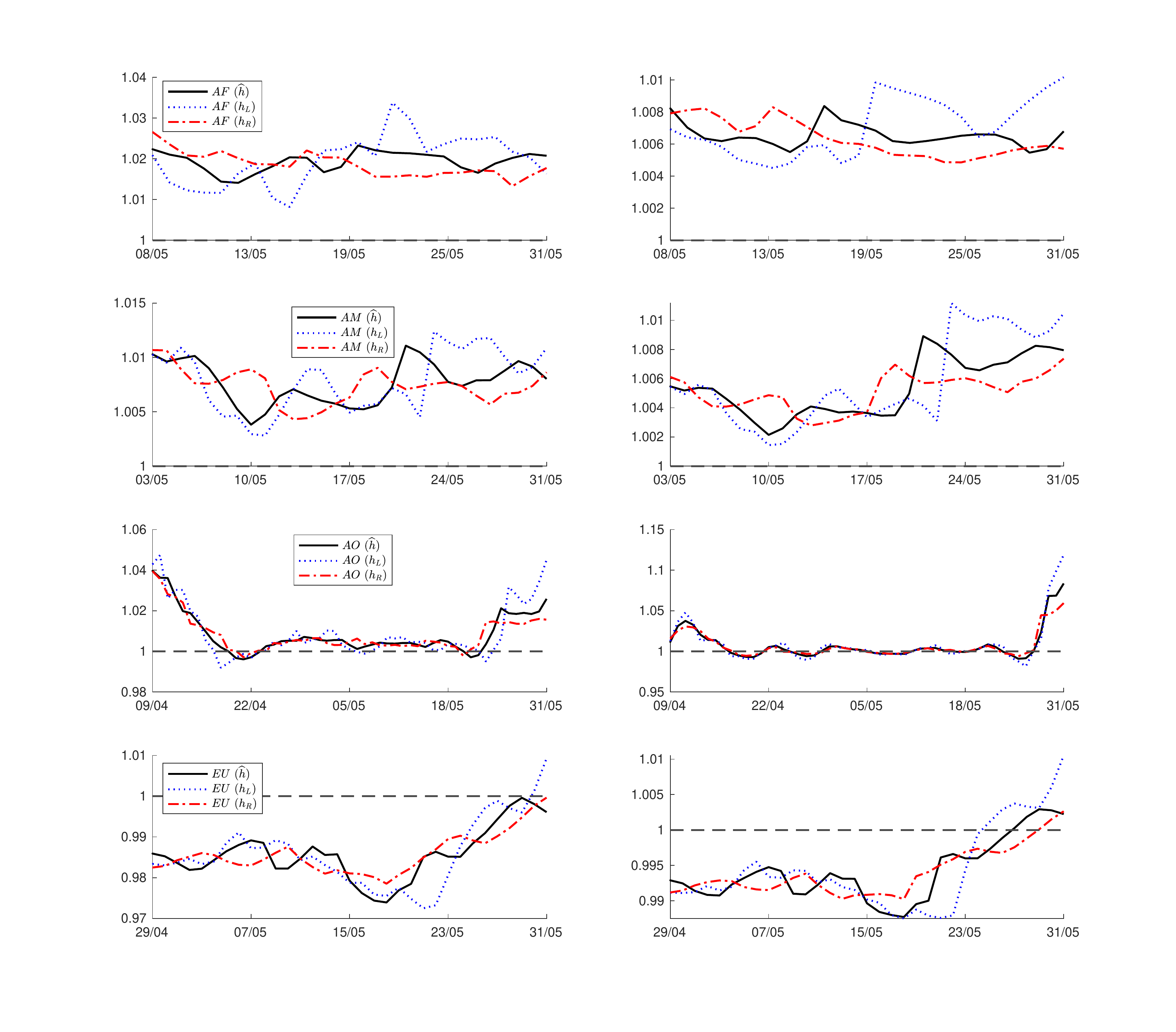} \\
\end{figure}

\newpage

\begin{landscape}
%\newgeometry{width=24cm, left=1cm, top=3cm}
\begin{figure}[H]\caption{Model 1 --- $Q_{\tau_t,i1}$ of Infection Data. The top and bottom panels are Case 1 and Case 2 respectively. The reference countries are presented in Table \ref{Table3}.}\label{Infection_Q}
\centering
\hspace*{0cm}\includegraphics[scale=0.53]{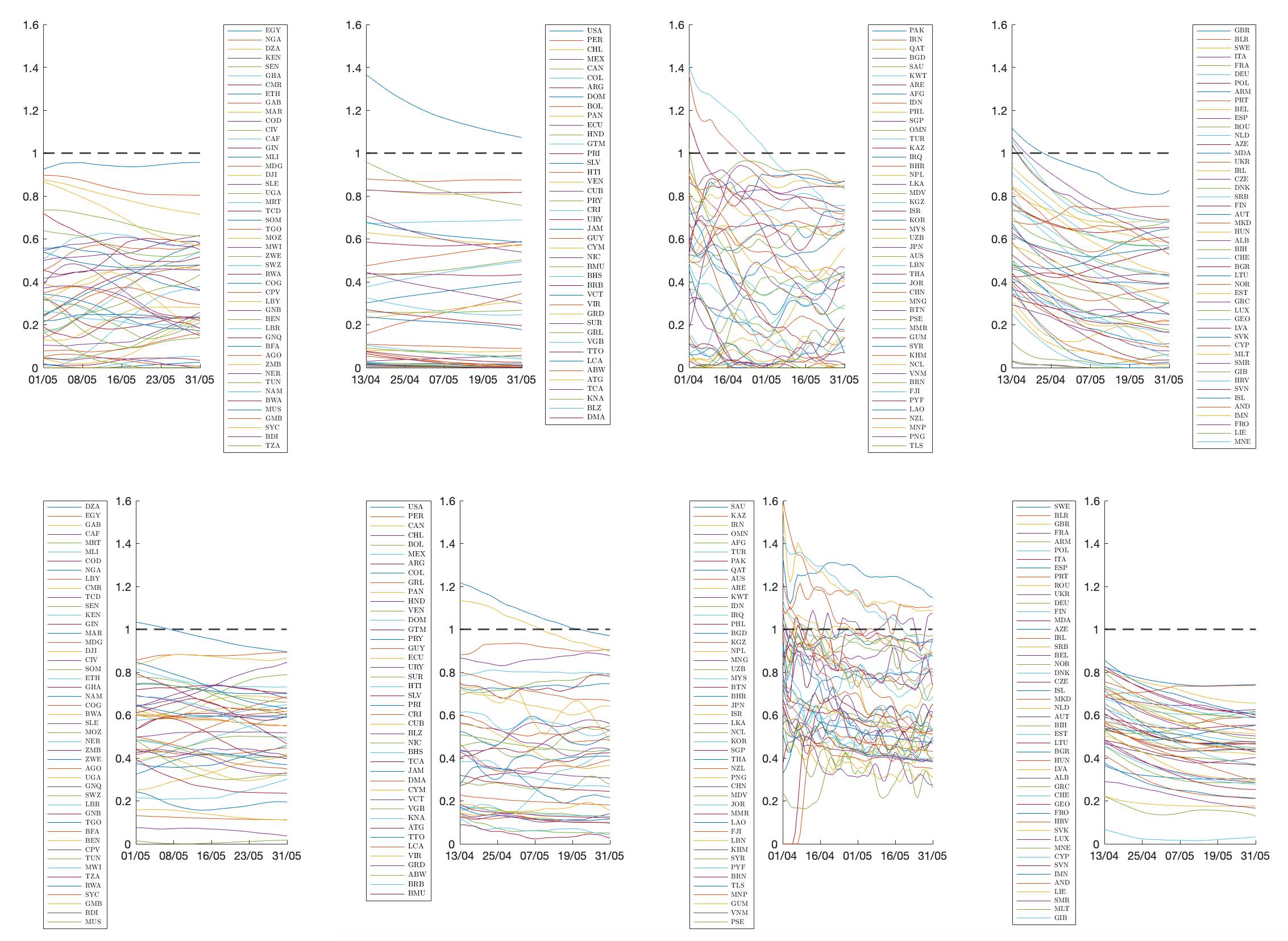} \\
\end{figure}
\end{landscape}

\begin{landscape}
%\newgeometry{width=24cm, left=1cm, top=3cm}
\begin{figure}[H]\caption{Model 1 --- $Q_{\tau_t,i1}$ of Death Data. The top and bottom panels are Case 1 and Case 2 respectively. The reference countries are presented in Table \ref{Table3}.}\label{Death_Q}
\centering
\hspace*{0cm} \includegraphics[scale=0.55]{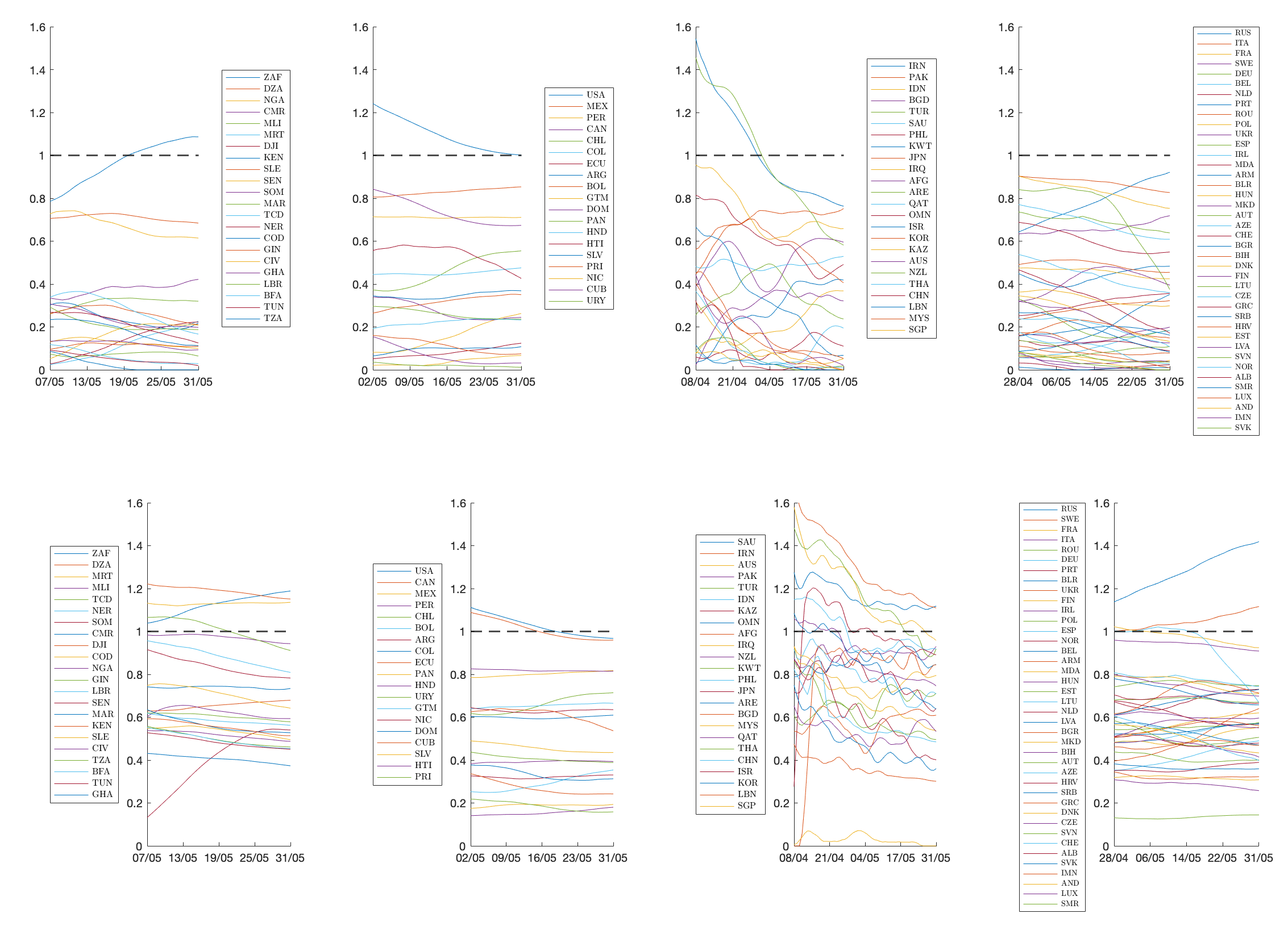}
\end{figure}
\end{landscape}

\begin{figure}[]\caption{Model 1 --- Estimated $a$ of Infection Data using Rolling Window. The left and right panels are Case 1 and Case 2 respectively.}\label{RW_a_Inf}
\centering
\hspace*{-1cm} \includegraphics[scale=0.5]{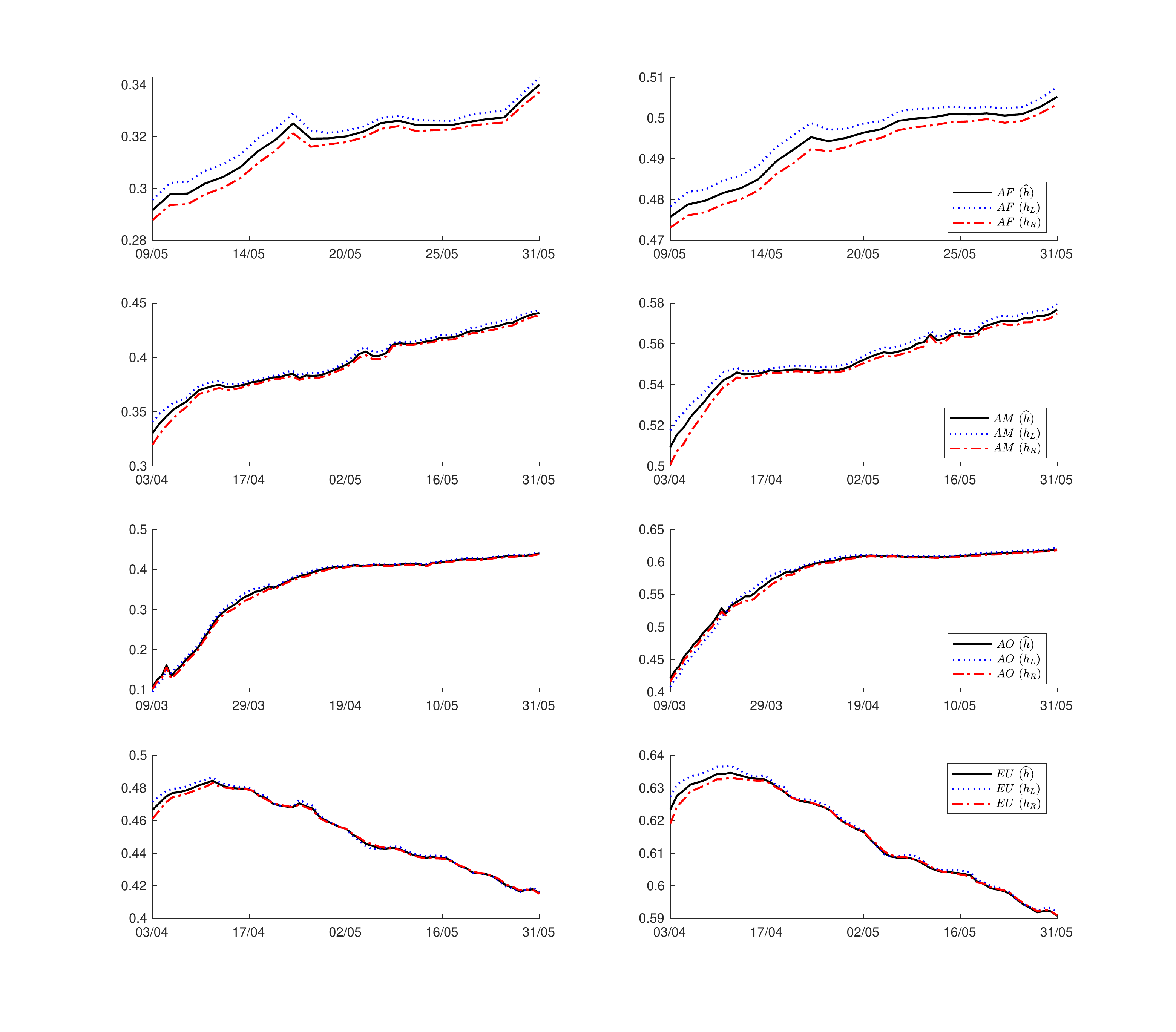} \\
\end{figure}

\begin{figure}[h]\caption{Model 1 --- Estimated $a$ of Death Data using Rolling Window. The left and right panels are Case 1 and Case 2 respectively.}\label{RW_a_Dea}
\centering
\hspace*{-1cm} \includegraphics[scale=0.5]{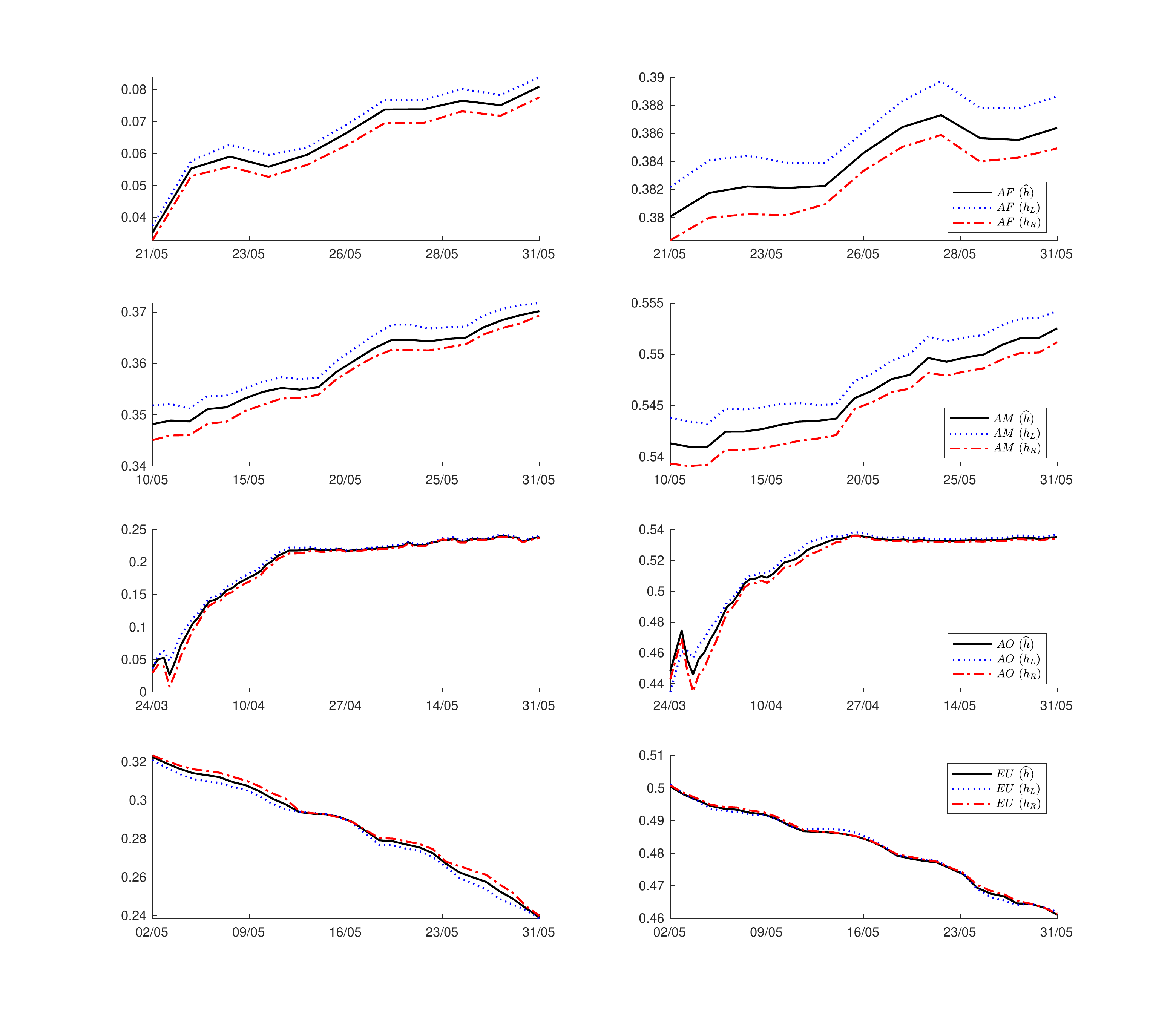} \\
\end{figure}

\begin{figure}[h]\caption{Model 1 --- $\bar{R}$ of Infection Data using Rolling Window. The left and right panels are Case 1 and Case 2 respectively.}\label{RW_R_Inf}
\centering
\hspace*{-1cm} \includegraphics[scale=0.5]{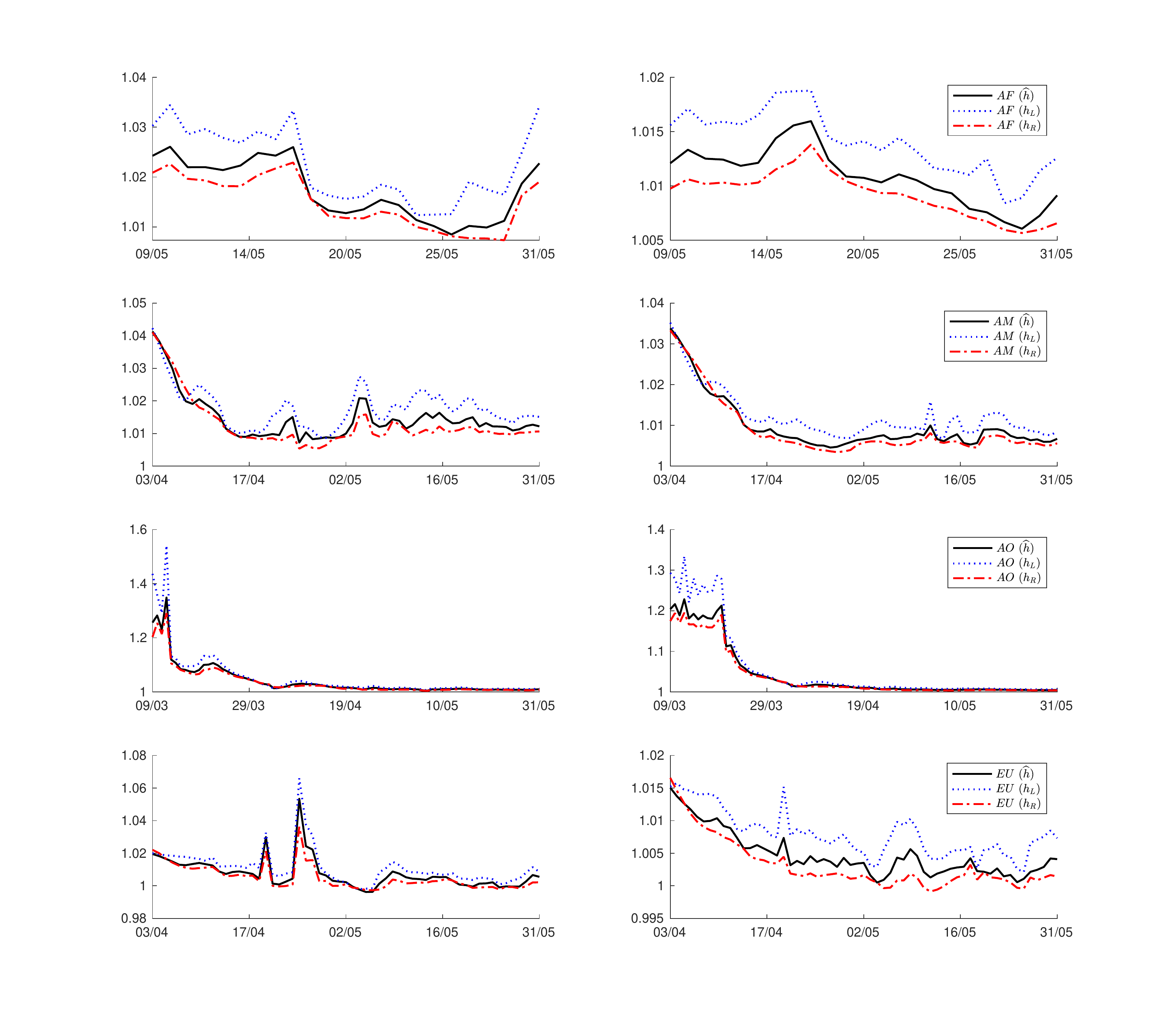} \\
\end{figure}

\begin{figure}[h]\caption{Model 1 --- $\bar{R}$ of Death Data using Rolling Window. The left and right panels are Case 1 and Case 2 respectively.}\label{RW_R_Dea}
\centering
\hspace*{-1cm} \includegraphics[scale=0.5]{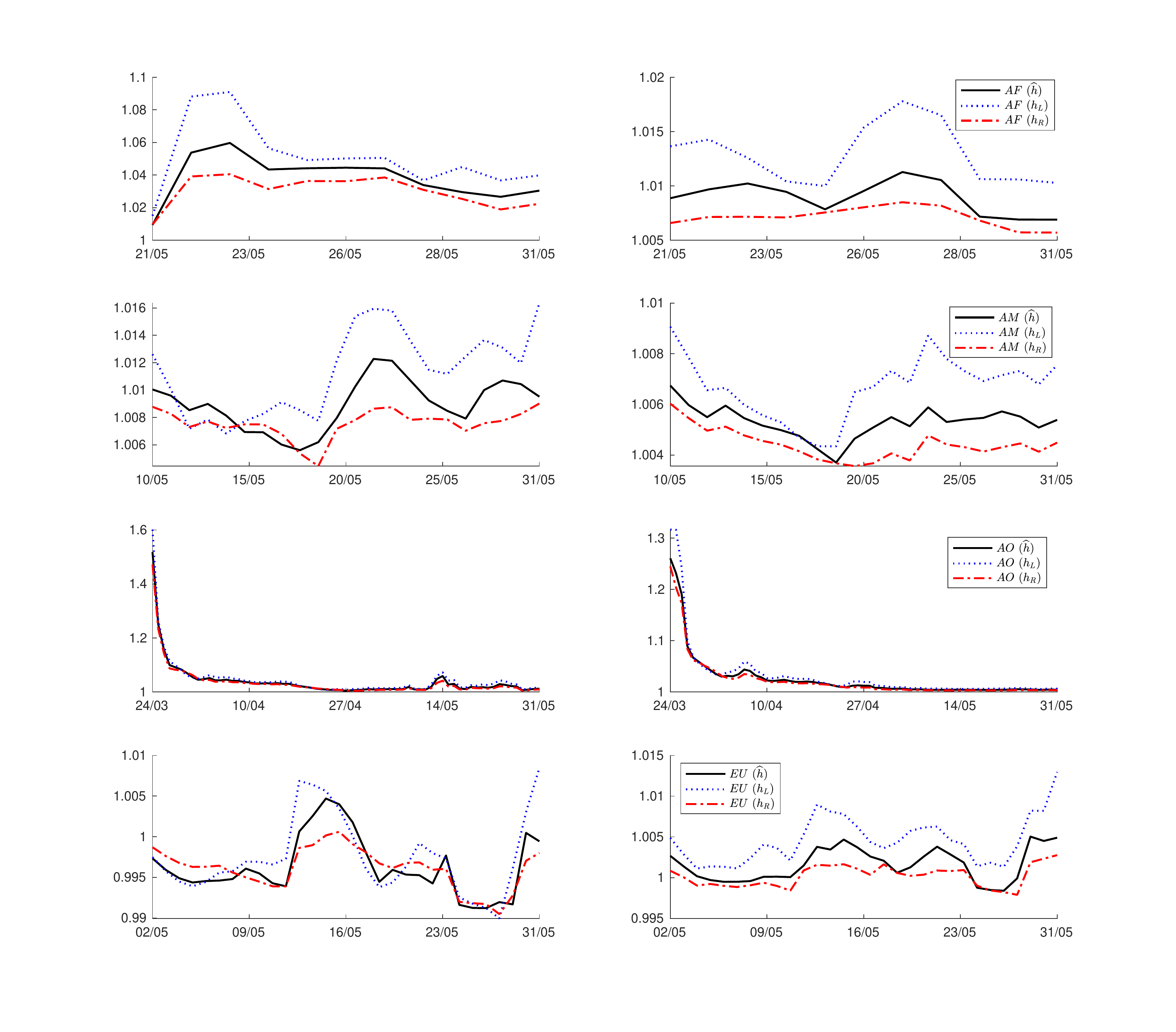} \\
\end{figure}

\begin{figure}[h]\caption{Model 2 --- $R_{t+1,t}$ of Infection Data. The left and right panels are Case 1 and Case 2 respectively. }\label{Infection_R_Extension}
\centering
\hspace*{-1cm} \includegraphics[scale=0.5]{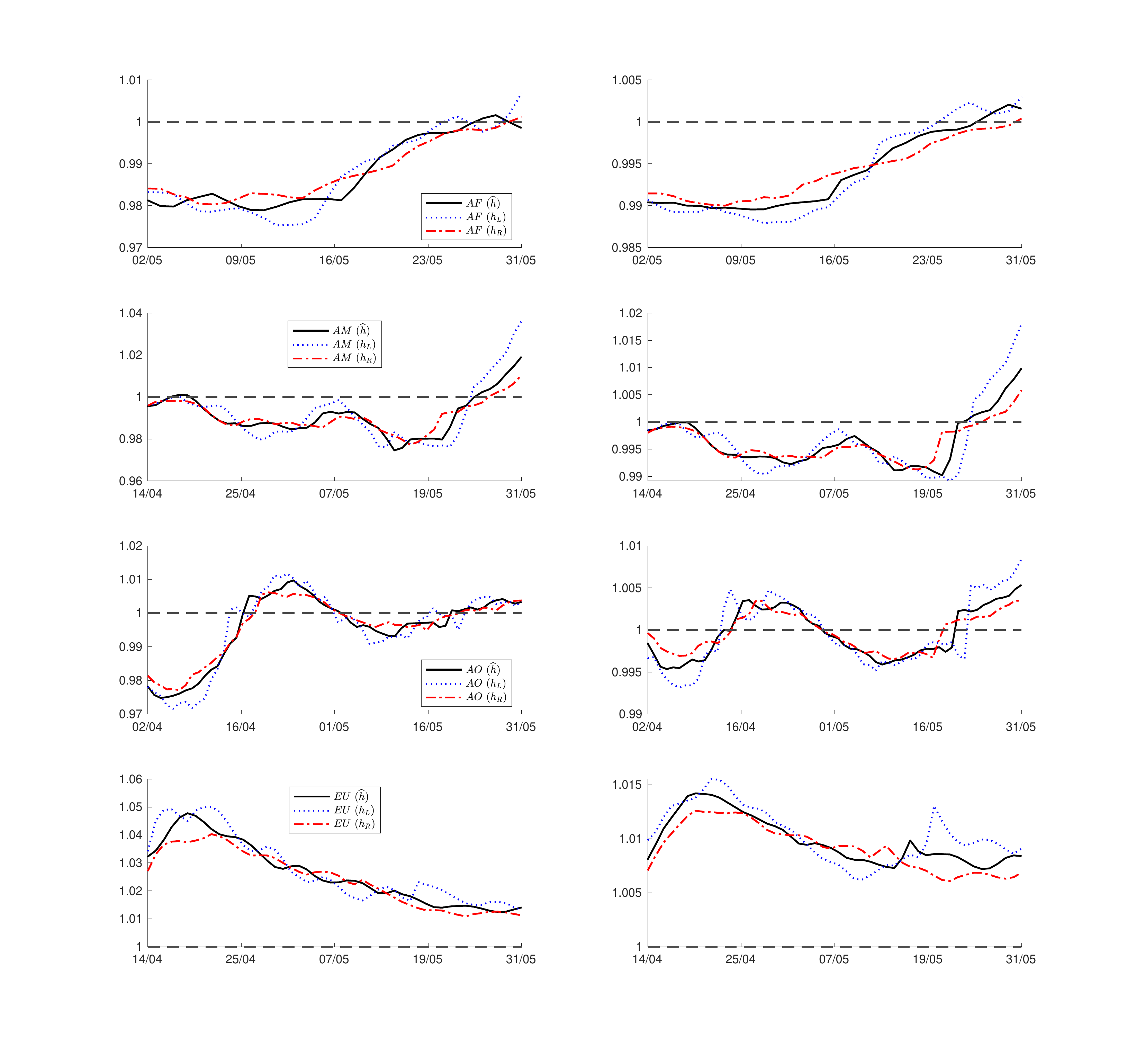} \\
\end{figure}

\begin{figure}[h]\caption{Model 2 --- $R_{t+1,t}$ of Death Data. The left and right panels are Case 1 and Case 2 respectively.}\label{Death_R_Extension}
\centering
\hspace*{-1cm} \includegraphics[scale=0.5]{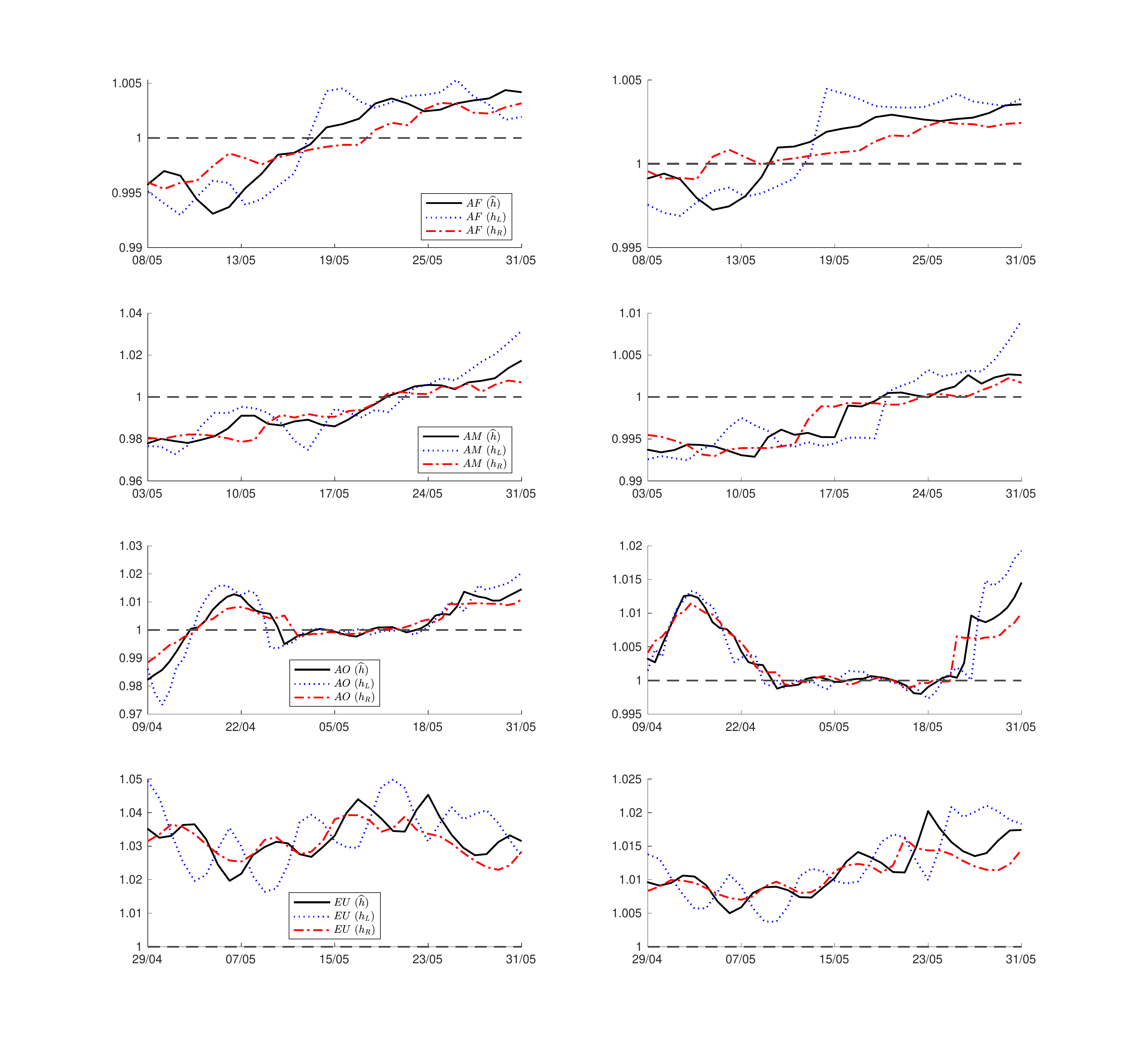} \\
\end{figure}

\begin{landscape}
%\newgeometry{width=22cm, left=1cm, top=3cm}
\begin{figure}[H]\caption{Model 2 --- $Q_{\tau_t,i1}$ of Infection Data. The top and bottom panels are Case 1 and Case 2 respectively. The reference countries are presented in Table \ref{Table3}.}\label{Infection_Q_Extension}
\centering
\hspace*{0cm}\includegraphics[scale=0.53]{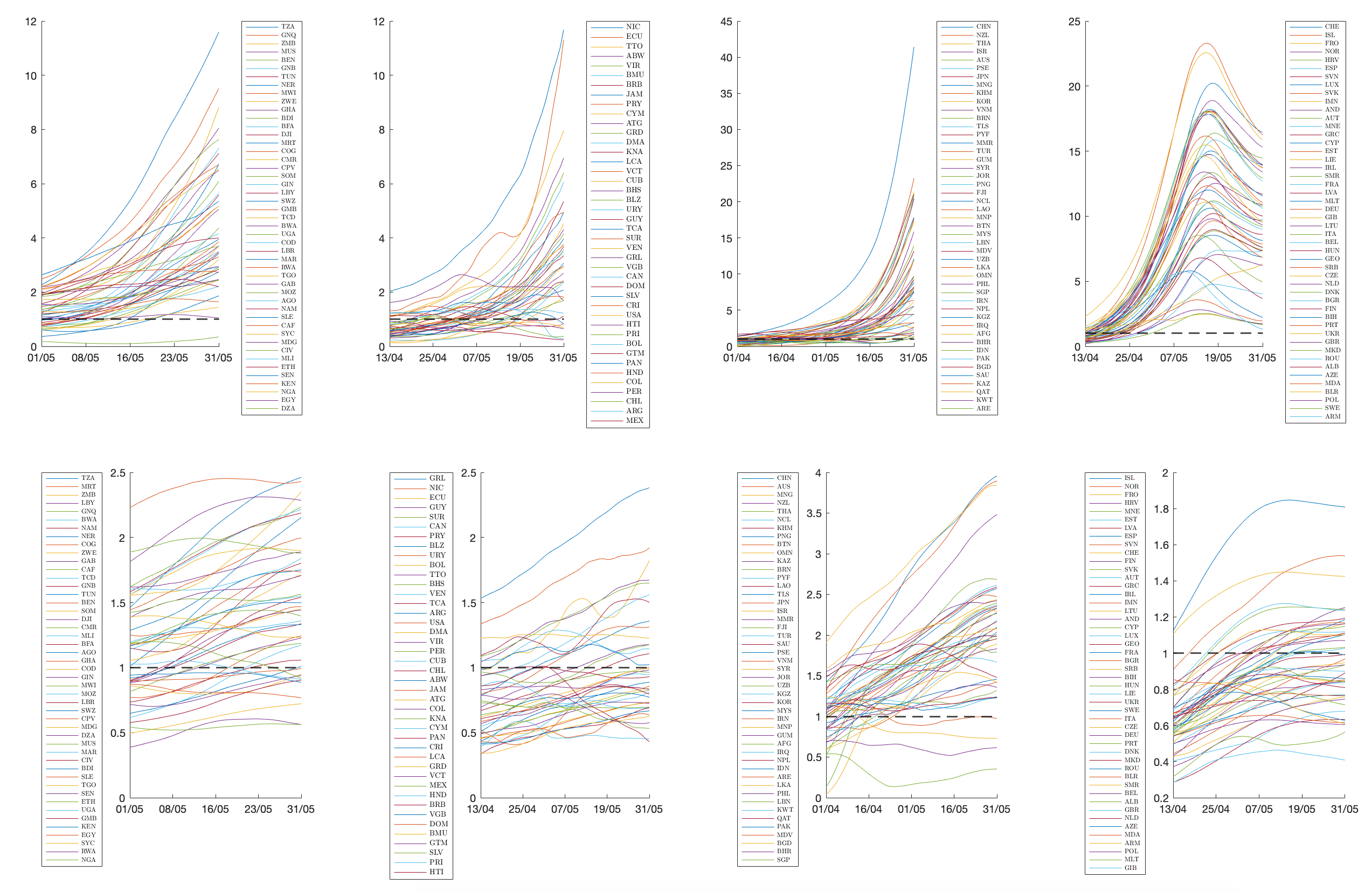} \\
\end{figure}
\end{landscape}

\begin{landscape}
%\newgeometry{ left=1cm, top=1cm, bottom = 1cm}
\begin{figure}[H]\caption{Model 2 --- $Q_{\tau_t,i1}$ of Death Data. The top and bottom panels are Case 1 and Case 2 respectively. The reference countries are presented in Table \ref{Table3}.}\label{Death_Q_Extension}
\centering
\hspace*{0cm} \includegraphics[scale=0.55]{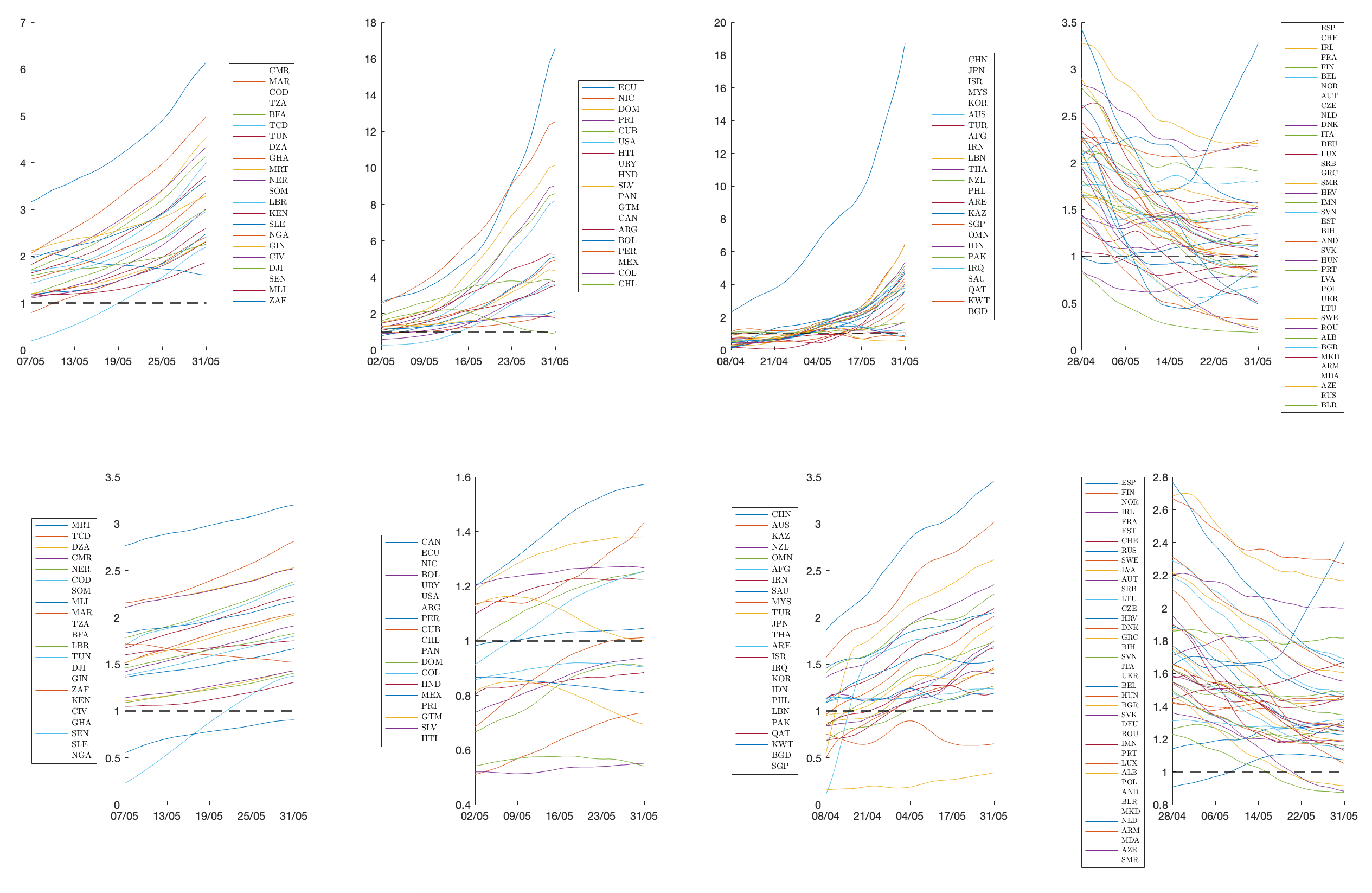} \\
\end{figure}
\end{landscape}

}
\end{document}